\documentclass[twocolumn]{article}
\usepackage{dcolumn}
\usepackage{bm}

\usepackage{graphicx}
\usepackage{bmpsize}

\usepackage[margin=25truemm]{geometry}

\title{Non-ergodic Phase Transition in the Global Hysteresis of the Frustrated Magnet $\mathrm{DyRu_2Si_2}$}

\author{S. Yoshimoto, Y. Tabata, T. Waki, and H. Nakamura \\
\textit{Department of Materials Science and Engineering, Kyoto University, Kyoto 606-8501, Japan} }
\date{}

\makeatletter
\newcommand{\tblcaption}[1]{\def\@captype{table}\caption{#1}}
\makeatother

\begin{document}

\maketitle

\begin{abstract} 
Some frustrated magnets  exhibit a huge hysteresis called ``global hysteresis (GH)'', where the magnetic plateaus appearing in the increasing field process are skipped in the decreasing field process from the high magnetic field state.
	In this paper, we focused on the frustrated magnet 
	$\mathrm{DyRu_2Si_2}$ and measured magnetization relaxations from two plateau states inside the GH loop, the phases III and IV, and investigated the phase transitions into them.
As a result of the relaxation measurements, no relaxation is observed in the phase III,
whereas long-time relaxations of more than $10^5$ sec are observed at the phase IV plateau.
Moreover, a Mpemba-effect-like relaxation phenomenon 
where the relaxation from an initial state prepared in the zero-field-cooled condition overtakes that from an initial state prepared in the field-cooled condition is observed.
These results indicate that the phase IV is the non-ergodic state with a complex free-energy landscape 
with multiple local minima, while the phase III has a simple free energy structure.
Therefore, the III-IV phase transition is considered to be the ergodic to non-ergodic phase transition.
Although this type of phase transition typically occurs in random glassy systems, 
the phase IV in
$\mathrm{DyRu_2Si_2}$ has a regular long-range ordered magnetic structure and yet exhibits non-ergodic properties, which is highly nontrivial.
Our findings open the possibility of observing non-ergodic states in frustrated magnets with regular long-range orders.

\end{abstract}

\section{Introduction}

\begin{figure}[t]
		\vspace{20pt}
		\centering
		\includegraphics[width=0.9\columnwidth]{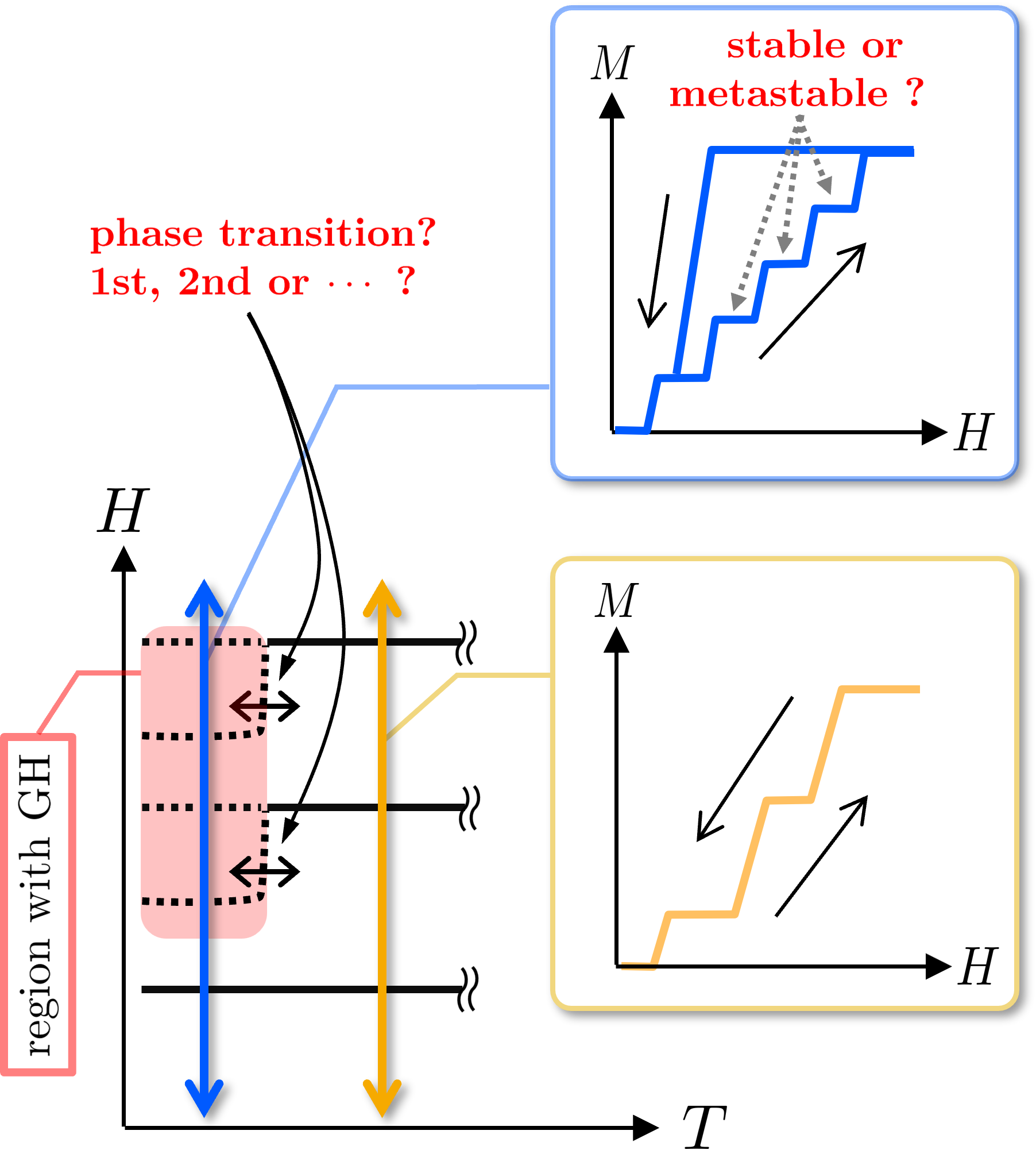}
		\caption{(Color online) Schematic picture of a phase diagram with a global hysteresis (GH) loop.
		 The GH loop usually appears in low temperatures and disappears with heating as shown in the panels with blue and yellow frames.
		The painted area indicates the GH region.
		In the phase diagram, the solid lines represent the phase transition lines, 
		whereas, the dotted lines inside the GH region correspond to 
		the magnetization jumps only observed in the IF process.
		 }
		\label{GlobHysSchem}
	\end{figure}

	In frustrated magnets, high degeneracy of the ground states due to the competition of magnetic interactions 
	causes a rich variety of exotic phenomena.
	Some of them exhibit multistep phase transitions against magnetic field and several magnetic plateaus are observed in their magnetization curves.
	One accepted understanding of this type of multistep phase transition is that, 
	when the nearly degenerated states exist at zero field, 
	the ground state of the system alternates with changing the magnetic field.
	Among such systems,  a huge hysteresis loop against magnetic field as schematically shown in Fig.
	\ref{GlobHysSchem} can be observed, where multiple plateaus appearing in the increasing-field (IF) process are skipped 
	in the decreasing-field (DF) process and the high-field phase directly transitions into the low-field phase
	\cite{J.Phys.Cond.7.1889,
	PhysicaB.212.343,
			Phys.Rev.B.97.134425,
			JPSJ.66.3996}.
	Such a huge hysteresis is referred to as ``global hysteresis (GH)'' hereafter. 
	In the GH loop, the high field state remains metastable at the lower fields.
	This is thought to result from strong nonequilibrium effects, indicating high energy barriers 
	between the metastable states and the plateau states observed in the IF process.

	The plateaus inside the GH loop are potentially interesting stages for studying metastability and phase transition.
	It is not always guaranteed that the plateau states inside the GH loop are thermal equilibrium states.
	This is because these plateaus are not selected in the DF process.
	Therefore, it is nontrivial whether entering these magnetization plateau states is a thermodynamic phase transition or not.
	Even if it is, what type of phase transition it is remains an open question.
	Also, regarding the temperature change, the GH loop is expected to disappear with increasing temperature as shown in Fig. \ref{GlobHysSchem}. 
	Regarding this point, the specifics, such as whether the phase transitions along with such disappearance or appearance of the GH loop with the temperature change occurs or not, 
	 are quite unknown.

	 The GH loop is observed in several materials such as
	 $\mathrm{DyRu_2Si_2}$
	 \cite{J.Phys.Cond.7.1889},
	 $\mathrm{TbRu_2}{X_2}$
	 ($X=$ Si or Ge)
	 \cite{PhysicaB.212.343},
	 $\mathrm{Tb_3Ni}$
	 \cite{Phys.Rev.B.97.134425} and
	 $\mathrm{Ca_3Co_2O_6}$
	 \cite{JPSJ.66.3996}.
	For example, $\mathrm{Ca_3Co_2O_6}$ exhibits the multistep metamagnetic transitions in low
temperature and 
	its magnetization hysteresis does not close over a wide range of magnetic field. 			
	Interestingly, strong nonequilibrium phenomena such as long-time relaxation from the magnetic plateaus are observed
	\cite{JPSJ.80.034701.,Phys.Rev.B.70.064424}.
	However, the cause of such phenomena is unknown and
	the above-mentioned questions regarding the changes to the states inside the GH loop have never been discussed in detail.
	Such intriguing nonequilibrium phenomena can be found in the GH loop and transitions into such a state worth studying.
	
	In this work, the GH loop of the frustrated magnet
	$\mathrm{DyRu_2Si_2}$ is investigated using a single crystal sample.
	This material belongs to the series of the intermetallic compounds
	$RT_2X_2$($R$ rare earth, $T$ 3d or 4d metal, $X$ Si or Ge),
	whose crystal structure is the
	$\mathrm{ThCr_2Si_2}$-type
	(space group:
	$I4/mmm$).
	The Dy$^{3+}$ ion has a magnetic moment 
	($\mu = 10\ \mu_\mathrm{B}$ for a free ion)
	and a strong uniaxial magnetic anisotropy along the c-axis.
	The 4f electrons of Dy$^{3+}$ are highly localized and the Dy spins interact through the oscillating long-range RKKY interactions,
	which are responsible for the frustration effect in this compound.
	Due to the frustration, 
	$\mathrm{DyRu_2Si_2}$ exhibits multiple phase transitions against temperature
	($T$) and magnetic field
	($H$).
	The $H \mathchar`- T$ phase diagram consists of 4 different  antiferromagnetic or ferrimagnetic phases, I, II, III, and IV
	\cite{J.Phys.Cond.7.1889}, where the period of each magnetic unit cell is 9 times that of the crystallographic unit cell along the a- or b-axis
	\cite{PhysicaB.346-347.99}.
	In this compound, a GH loop was observed at 1.5 K in the previous research
	\cite{J.Phys.Cond.7.1889}.
	In the IF process, the magnetic plateaus corresponding to the phases II, I, III, and IV appear in sequence and 
	the forced ferromagnetic (FFM) state is reached at around the external field of 25 kOe.
	On the other hand, in the DF process, the FFM state directly transitions into the phase I, skipping the plateaus of the phases III and IV.
	Since the phase III and IV plateaus are inside the GH loop, it is nontrivial if these plateau states are equilibrium or not, 
	and details on the transition to these phases, III and IV, against magnetic field and temperature are unknown.

	To elucidate these issues, we conducted detailed measurements of the magnetization hysteresis and magnetization relaxations.
	As a result, long-time relaxations exceeding $10^5$ sec are observed in the phase IV plateaus, while no relaxation is observed in the phase III plateau.
	Moreover, in the phase IV, the relaxation measured in the zero-field-cooling (ZFC) condition overtakes that measured in the field-cooling (FC) condition.
	This phenomenon resembling the Mpemba effect is very noteworthy
	\cite{PhysEdu.4.1969}.
	The results indicate that the phase IV plateau is metastable, whereas the phase III plateau is stable.
	It is also indicated that the phase IV is a non-ergodic phase with a complicated multi-valley free energy landscape, 
	where multiple local minima are separated from each other with high energy barriers.
	On the other hand, the phase III has a simple free-energy landscape.
	Thus, the phase transition into the phase IV inside the GH loop, which is the III-IV  phase transition in this case, is the ergodic to non-ergodic phase transition.

\section{Experimental Methods}
\label{exp}

	Single crystal of $\mathrm{DyRu_2Si_2}$ was grown by the Czochralski method with a tetra-arc furnace.
	It was then cut parallel to the c-plane and 10.1 mg of a cubic-like shaped sample was used for magnetization measurements.
	The measurements were performed using the SQUID magnetometer (MPMS, Quantum Design) 
	equipped in the Research Center for Low Temperature and Materials
Sciences, Kyoto University, in the temperature range of 1.8-50 K and field range of 0-30 kOe.
	Throughout all the measurements, magnetic field was applied parallel to the c-axis.
	The demagnetization effect is considered by approximating the sample shape as a sphere.
	Therefore, the demagnetization field coefficient is
	$4\pi/3$ in cgs unit and the effective field is given as
	\begin{equation}
		H_\mathrm{eff} 
		= 
		H_\mathrm{ex} - 
		\frac
		{4 \pi}
		{3} 
		M ,
	\end{equation}
	\noindent
	where 
	$H_\mathrm{ex}$ and $M$ are the external field and magnetization, respectively.
	To investigate the temperature change of the GH loop, the isothermal magnetization curves on both the IF and DF processes up to 
	$H_\mathrm{ex}=30$ kOe were measured at each temperature ranging from 2 to 15 K.
	Also at 2 K, several minor loops in the GH loop were measured. 
	Before every measurement of the magnetization curve and the minor loop, the sample was cooled in the ZFC condition from 50 K, where the system is paramagnetic, 
	to erase some possible memory effects at lower temperatures.
	Magnetization relaxations were measured from the initial states at several temperatures and magnetic fields in the phases III and IV.
	The initial states were chosen plateau states and other metastable states in the GH loops.
	Before every relaxation measurement, the sample was cooled from 50 K as well.
	Details on the preparations of these initial states are described with each result in the next section and the Appendix.
	Regarding some relaxations in the phase IV, the magnetizations are normalized in the way described in the Appendix, which is important for strict and qualitative discussion.

\section{Results}

	\subsection{Magnetic characterization and $H \mathchar`- T$ phase diagram}
	\label{Mag_Char}

	\ \\
	\noindent 	\textbf{$\bullet$ Magnetic susceptibility }

\begin{figure}
		\centering
		\includegraphics[width=.9\columnwidth]{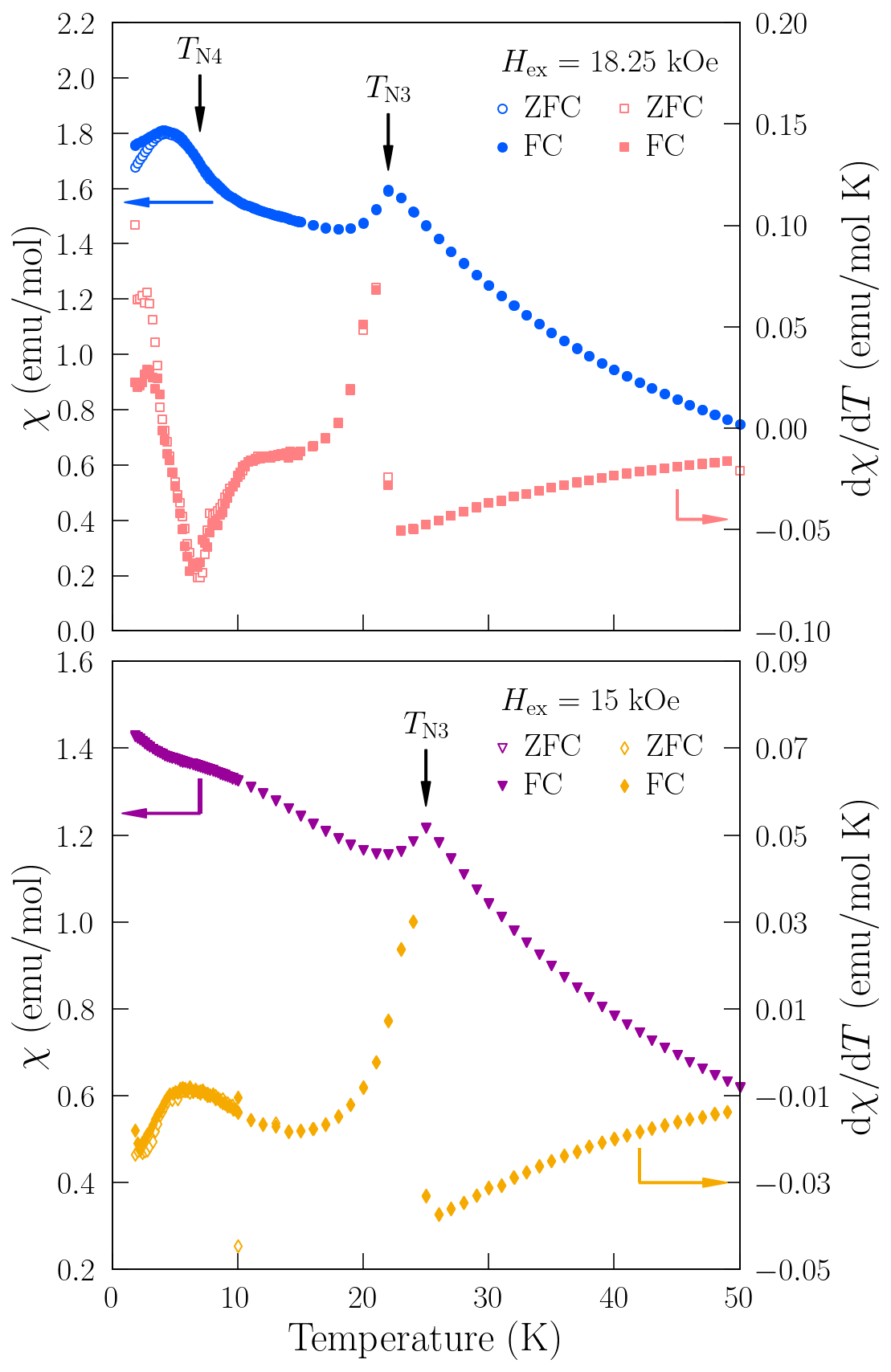}
		\caption{(Color online) Temperature dependences of susceptibilities and their temperature derivatives at 
					$ H_\mathrm{ex}= 18.25$ and 15
 kOe 
		measured on heating after ZFC or FC. 
		}
		\label{chidc}
	\end{figure}

	Figure
	\ref{chidc} is the temperature dependences of susceptibility and their temperature derivatives at 
	$H_\mathrm{ex} = 15$ and 18.25 kOe.
	These were measured on heating after cooling in the ZFC or FC conditions.
	At 15 kOe, little deviation between the ZFC and FC susceptibilities is seen.
	Each susceptibility exhibits a cusp anomaly at
	$T_\mathrm{N3} = 25\ \mathrm{K}$, corresponding to the para-III phase transition.
	The susceptibility shows a minimum just below $T_\mathrm{N3}$ and then monotonically increases with decreasing temperature.
	At 18.25 kOe,
	the cusp anomaly at $T_\mathrm{N3} = 22$ K and the minimum just below this temperature are seen as well.
	Notably, at this magnetic field, the temperature derivatives exhibit dip anomalies 
	at $T_\mathrm{N4} = 7.5\ \mathrm{K}$, indicating the III-IV phase transition in both cooling conditions.
	Moreover, at around 5 K, the susceptibilities exhibit broad peaks, and
	 deviate from each other below this temperature,
	indicating nonequilibrium effects in the phase IV.
	On the other hand, since little deviation is seen at 15 kOe, strong nonequilibrity is less expected in the phase III.
	The comparison in the nonequilibrium effects is extensively investigated in subsection 3.3.

	\ \\
	\noindent \textbf{$\bullet$ GH loop}
	
\begin{figure*}[p]
		\centering
		\begin{tabular}{cc}
		\begin{minipage}{0.45\linewidth}
			\centering
			\includegraphics[height=23cm]{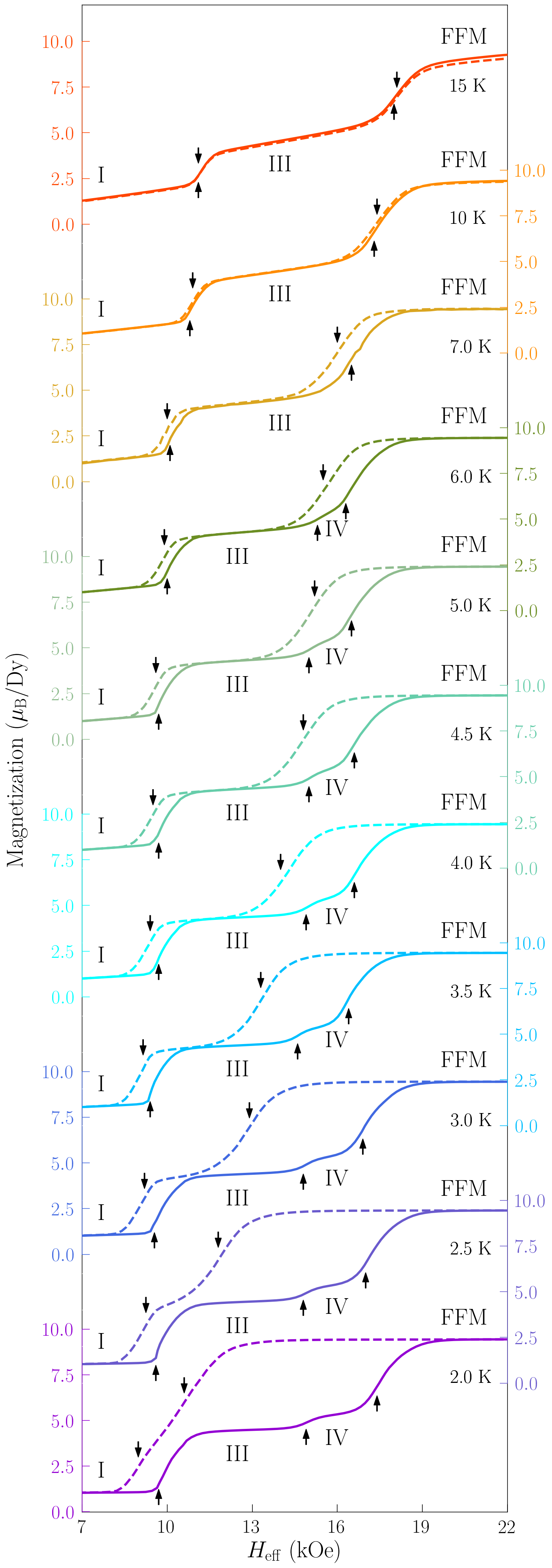}\\
			{\small (a) }
		\end{minipage}
		\hspace{0.5cm}
		\begin{minipage}{0.45\linewidth}
			\centering
			\includegraphics[height=23cm]{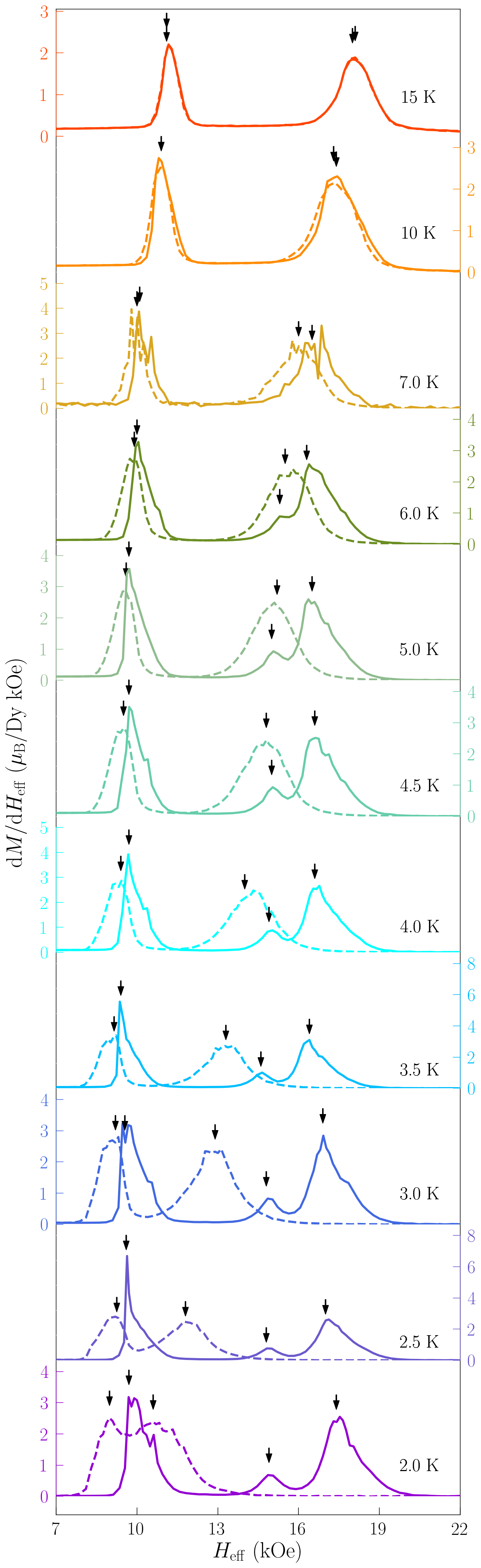}\\
			{\small (b) }
		\end{minipage}
		\end{tabular}
		\caption{(Color online) 
					(a) Magnetization hysteresis loops and 
					(b) the effective field derivatives of magnetization in the corresponding hysteresis loops.
					The solid and dashed curves represent the magnetizations measured in the IF and DF processes and the corresponding derivative curves, respectively.
					In the panel (a), plateaus appearing in the IF branches are labeled.
					In the panel (b), the arrows indicate the effective field giving the peak anomalies in each derivative curve and 
					those effective fields are also indicated on the corresponding magnetization curves in the panel (a).
					Note that the hysteresis loops were measured in the range of 	$0 \leq H_\mathrm{ex} \leq 30 \ \mathrm{kOe}$  but the data in the range of
					$7 \ \mathrm{kOe} \leq H_\mathrm{eff} \leq 22 \ \mathrm{kOe}$,
					where the GH loops are observed, is shown.
					}
		\label{M-H}
	\end{figure*}

	Figures
	\ref{M-H} (a) and (b) show the results of magnetization hysteresis measurements and
	their effective field derivatives.
	The solid and dashed curves represent the magnetization measured in the IF and DF processes, respectively.
	Note that the horizontal axis represents the effective field given by eq. (1).
	In the IF branch at 2 K, the phase I, III, and IV plateaus appear in sequence, as assigned in the literature, 
	and the magnetization saturates at around $H_\mathrm{eff} = 20\ \mathrm{kOe}$, which is the FFM state.
	On the other hand, in the DF branch, the magnetization does not return to the phase IV or III plateau appearing in the IF branch 
	and directly goes back to the phase I plateau, forming the GH loop, which reproduces the GH loop at 1.5 K in the previous work
	\cite{J.Phys.Cond.7.1889} as described in Sect. 1. 
	The peaks representing the phase transitions between these phases are also confirmed in the corresponding derivative curve (Fig. \ref{M-H} (b)).  
	In this derivative curve of the DF branch, one can see two broad and not fully separated peaks around 10 kOe,
	which indicates a possible phase transition into the phase III.
	However, since the clear magnetization plateau with the magnetization of the phase III is not observed in the DF branch, 
	we consider that the FFM state directly transitions into the phase I, skipping the phases III and IV.

	\begin{figure*}[t]
		\centering
		\vspace{1cm}
		\begin{tabular}{cc}
			\begin{minipage}{0.45\linewidth}
				\centering
				\includegraphics[width=6cm]{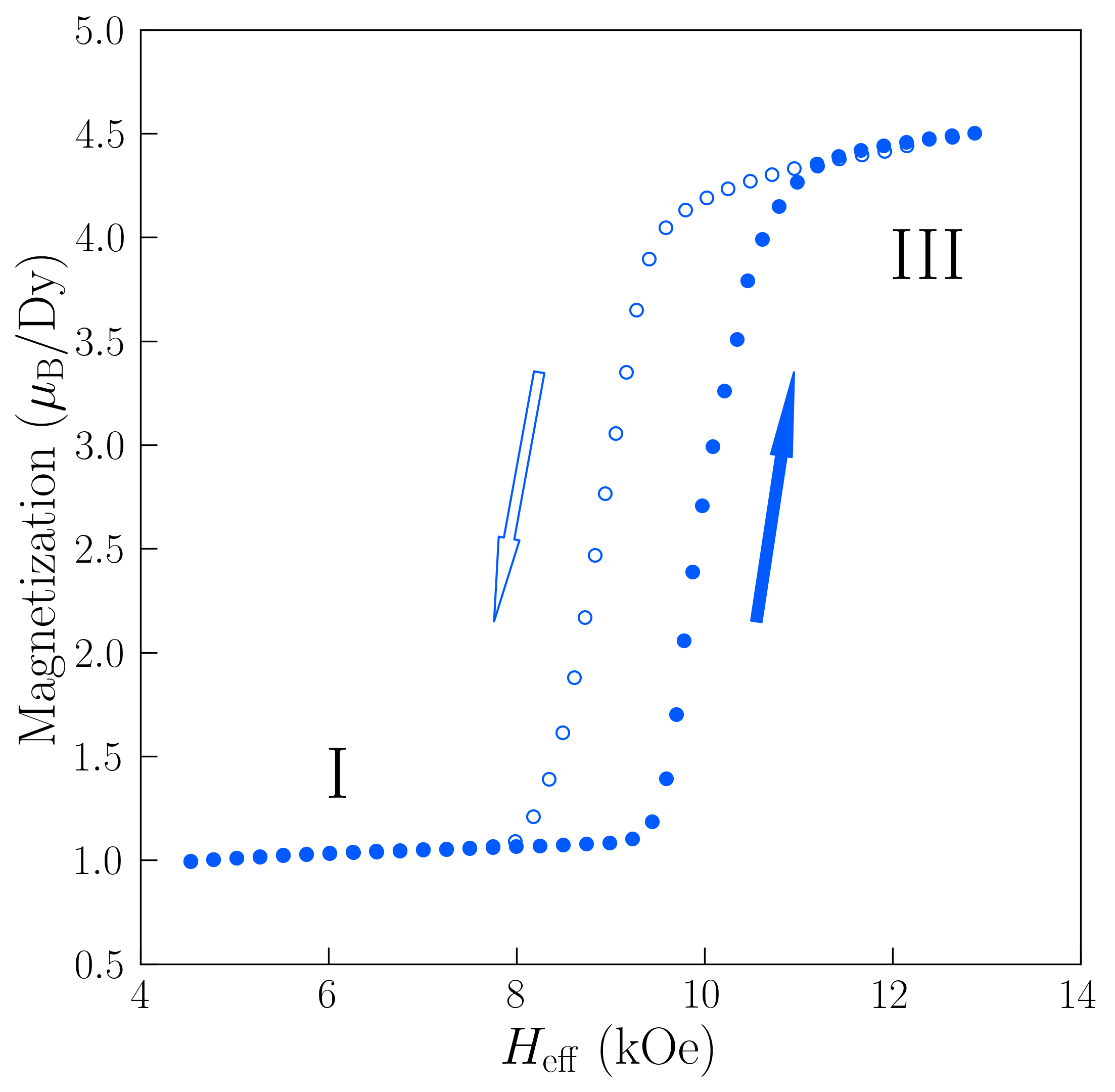}
				\\ 
				\centering {\small (a)}
				\label{small_minor_I-III}
			\end{minipage}
			\hspace{0cm}
			\begin{minipage}{0.45\linewidth}
				\centering
				\includegraphics[width=6cm]{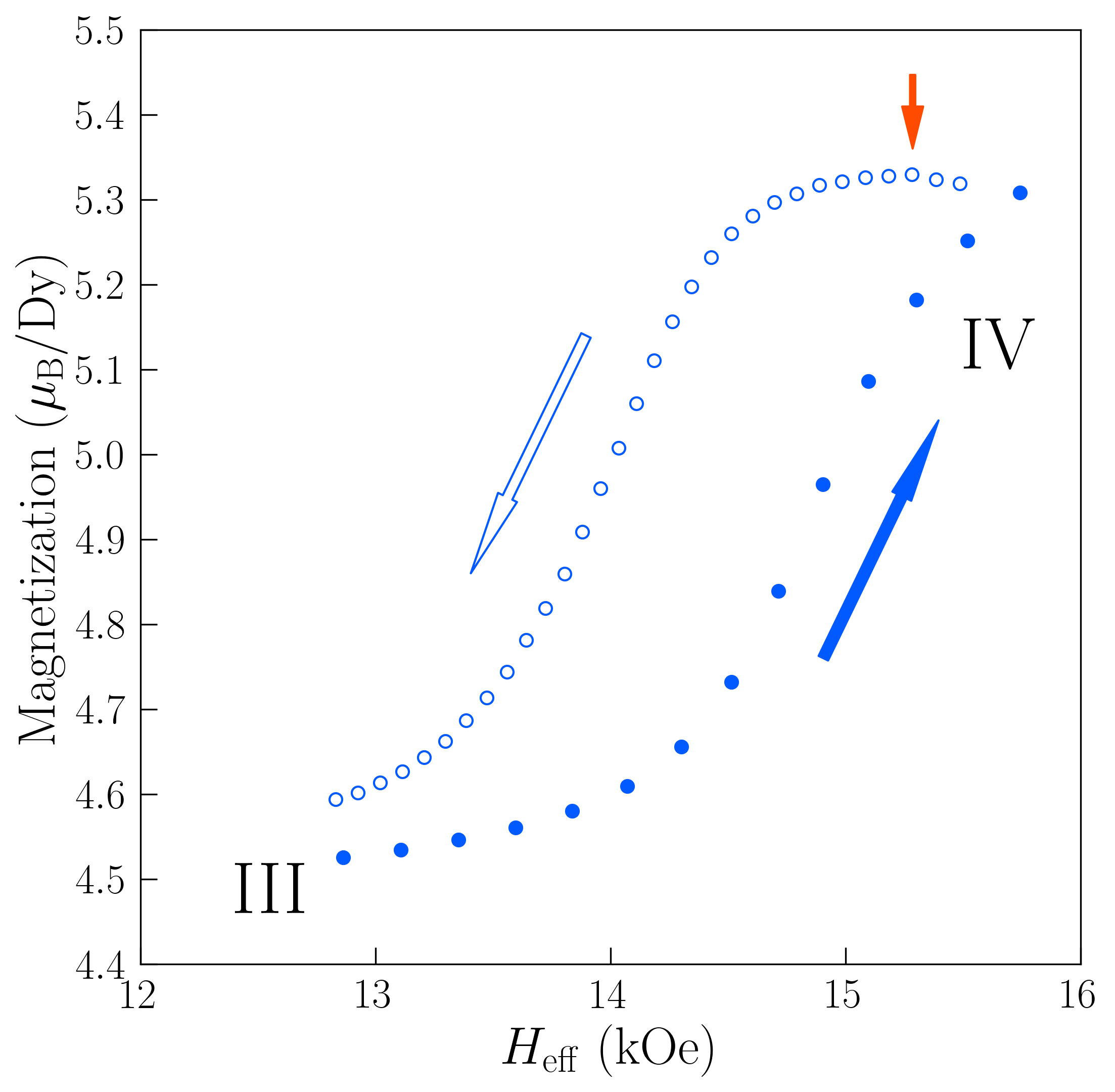}
				\\ 
				\centering {\small (b)}
				\label{small_minor_III-IV}
			\end{minipage}
		\end{tabular}
		\caption{(Color online) 
					Minor loops between
					(a) the phase I and III plateaus, and (b) the phase III and IV plateaus
					inside the GH loop at 2 K.
					The red arrow in the panel (b) indicates the maximum magnetization in the DF branch.}
		\label{small_minor}
	\end{figure*}

			\begin{figure*}[t]
				\centering
				\vspace{1cm}
				\begin{tabular}{cc}
					\begin{minipage}[b]{0.45\linewidth}
						\centering
						\includegraphics[height=6cm]{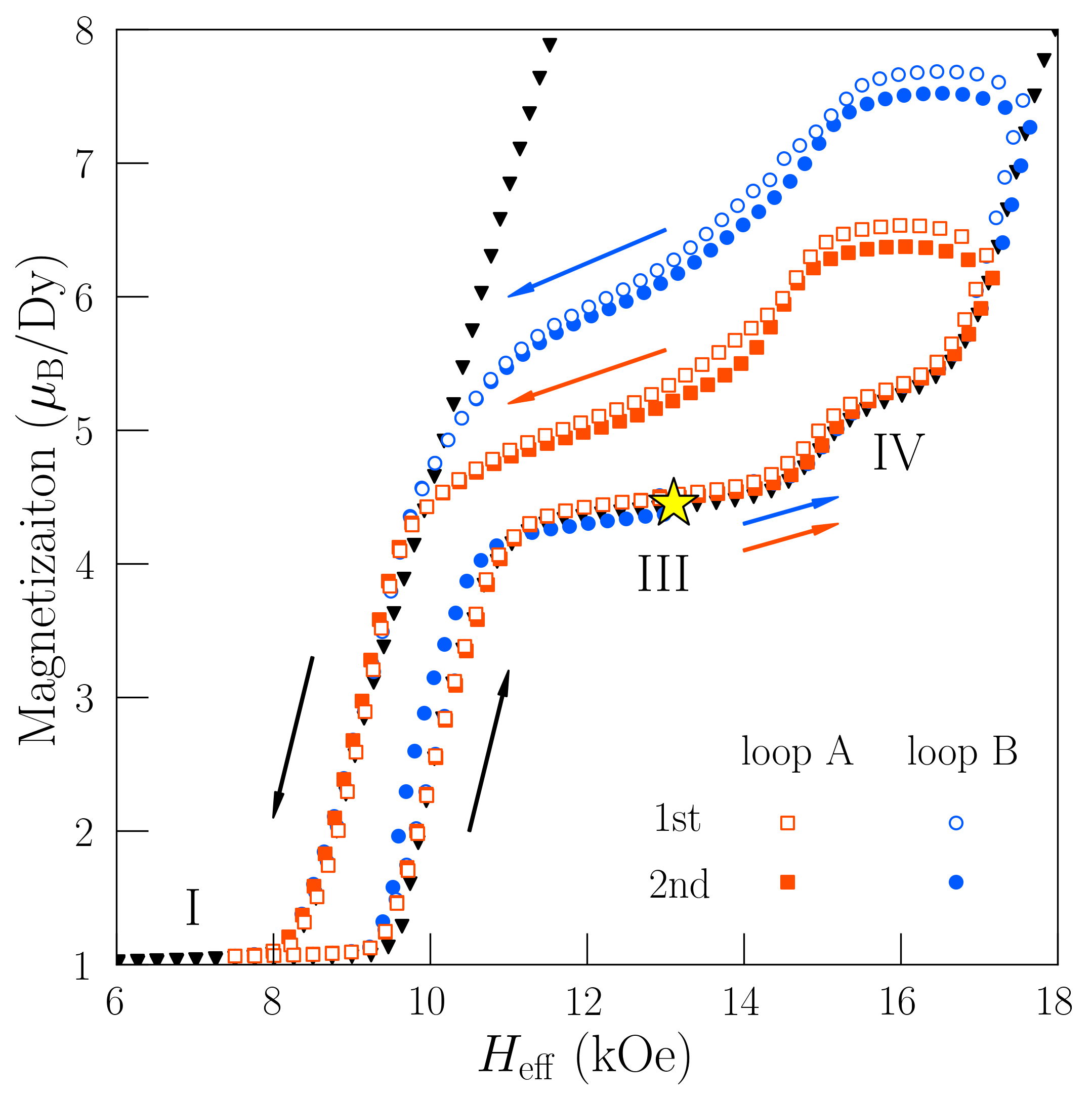}
						\\ 
						\centering {\small (a) }
					\end{minipage}
					
					\begin{minipage}[b]{0.45\linewidth}
						\centering
						\includegraphics[height=6cm]{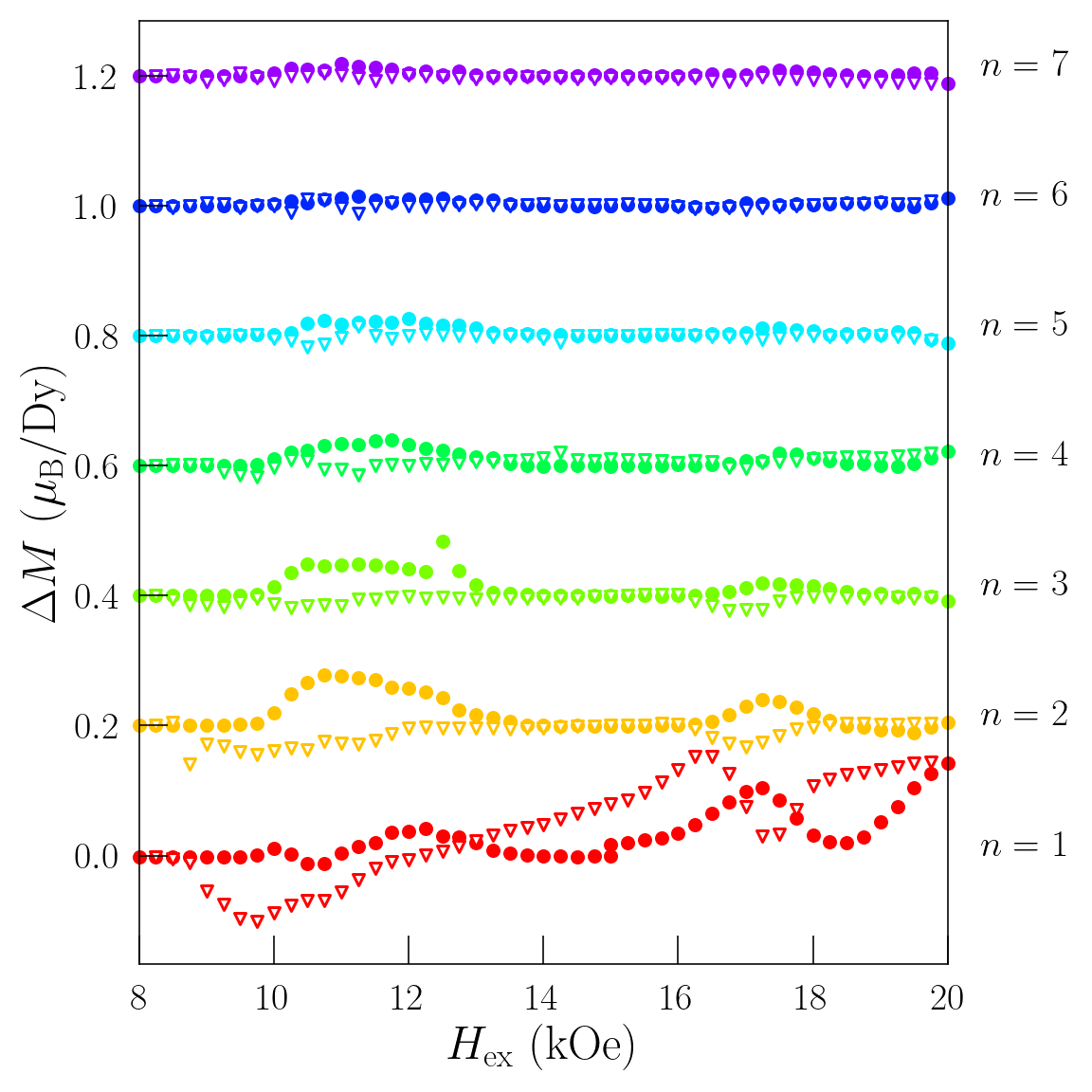}
						\\ 
						\centering {\small (b) }
					\end{minipage}
				\end{tabular}
				\caption{(Color online) 
								(a) Minor loops A ($8\ \mathrm{kOe} \leq H_\mathrm{ex} \leq 20\ \mathrm{kOe}$) and 
								B ($8\ \mathrm{kOe} \leq H_\mathrm{ex} \leq 21\ \mathrm{kOe}$) inside the GH loop at 2 K.
								The open and solid markers represent the 1st and 2nd cycles, respectively.
								The starting point of the loops is indicated with the yellow star.
								Black triangle markers represent the whole loop, i.e. the curve at 2 K in Fig. \ref{M-H} (a).
								(b) Magnetizations difference between the consecutive $n$th and $(n+1)$th cycles of the loop A.
								For example, the plot of $n=1$ represents the difference of the 1st and 2nd cycles.
								The solid circles and open triangles represent the difference in the IF and DF branches, respectively.
								The data are shifted sequentially for clarity.
								}
				\label{minor_loop_wh}
			\end{figure*}

	Along with the temperature increase, the widths of the GH loop and the phase IV plateau become narrower. 
	At 2.5 K, a plateau-like structure with the same magnetization as the phase III plateau appears in the DF branch,
	which is pronounced more clearly at 3.0 K.
	Along with this, the ill-separated peak at 10 kOe of the derivative curve of the DF branch at 2 K gets separated more
	and the peaks of the FFM-III and III-I phase transitions are clearly observed.
	At 3.5 K, the phase III plateau in the DF branch touches that in the IF branch and the whole GH loop splits into two loops.
	One is the first-order-phase-transition (FOPT)-like loop between the phase I and III plateaus, and 
	the other is the smaller GH loop in which
	the phase IV plateau in the IF branch is skipped in the DF process and the FFM state directly transitions into the phase III. 
	With further increasing temperature, the widths of each loop and the phase IV plateau get narrower.
	The phase IV plateau in the IF branch disappears at 7 K so does the peak of the III-IV phase transition on the corresponding derivative curve.
	Note that, in the DF processes, neither the phase IV plateau nor the peak along with the FFM-IV phase transition is observed at any temperature.
	In other words, the phase IV plateau is always inside the GH loop.
	Above 7 K, the smaller GH loop transitions to a FOPT-like hysteresis loop between the phases III and FFM plateaus.
	At 15 K, these two FOPT-like hysteresis loops corresponding to the I-III and III-FFM transitions disappear and these phase transitions become second-order-like.

	\ \\
	\noindent \textbf{$\bullet$ Minor loops and characterization of the type of the I-III and III-IV phase transition}

	The type of phase transition between the plateaus inside the GH loop was investigated 
	by measuring the minor loops between the phase I and III plateaus 
	(Fig. \ref{small_minor} (a)) and between the phase III and IV plateaus
	(Fig. \ref{small_minor} (b)).
	On these measurements, the sample was first cooled in the ZFC condition and 
	then 
	$H_\mathrm{ex} = 8$ or 15 kOe were applied 
	to reach the phase I or III plateau, respectively, starting from which the loops were measured.
	In Fig. \ref{small_minor} (a), the loop closes at around $H_\mathrm{eff}=8\ \mathrm{kOe}$ 
	in the phase I plateau, indicating that the I-III phase transition is first-order.
	On the other hand, as shown in 
	Fig. \ref{small_minor} (b), the hysteresis loop between the phase III and IV plateaus doesn't close,
	which makes it difficult to classify the III-IV phase transition and one can not expect a simple mechanism for this phase transition.
	Also, we note that the DF branch in 
	Fig. \ref{small_minor} (b) shows a broad peak at around 
	$H_\mathrm{eff}=15$ kOe as indicated with the red arrow, which is just below the maximum field.
	In general, magnetization decreases monotonically with decreasing magnetic field in a static process, so this DF process seems to be a nonequilibrium process and, thus, casts doubt on the thermal equilibrium of the phase IV plateau.

	\ \\
	\noindent \textbf{$\bullet$ Minor loop and metastable states inside the GH loop}

	\begin{figure*}[t]
		\centering
		\includegraphics[width=\linewidth]{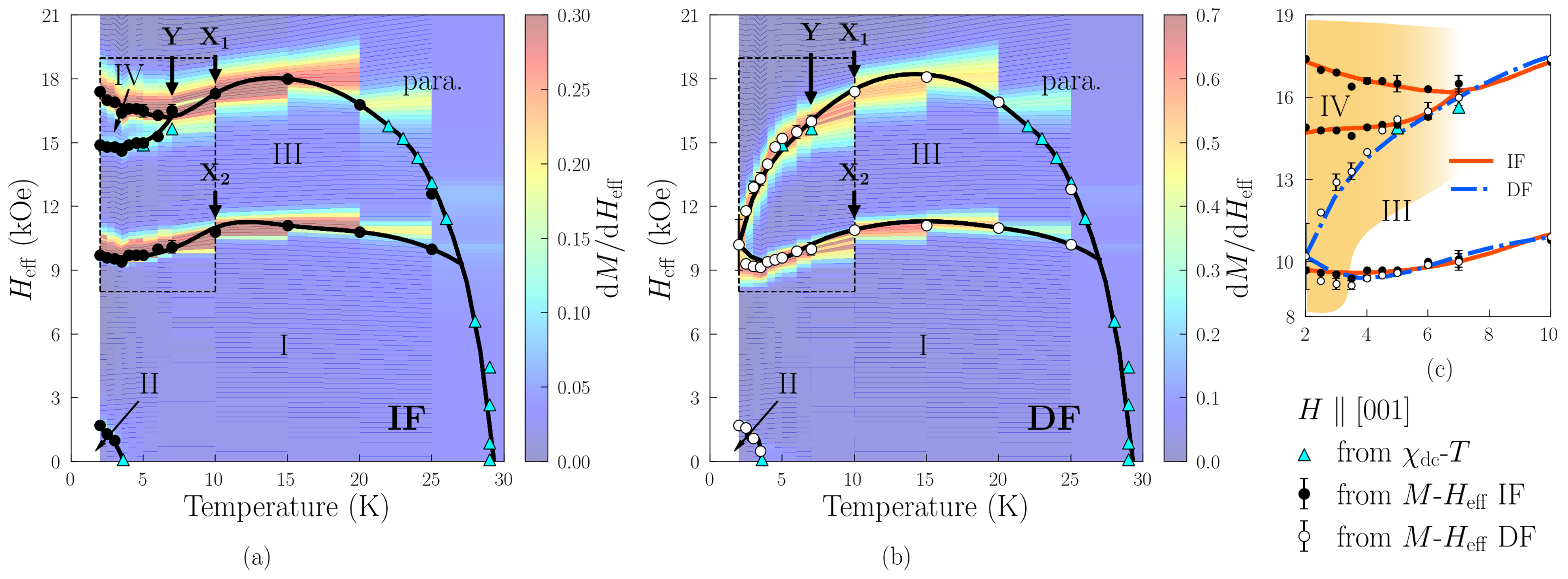}
		\caption{(Color online) 
						$H_\mathrm{eff} \mathchar`- T$ phase diagrams of
						$\mathrm{DyRu_2Si_2}$.
						The transition fields for the IF and DF branches of the $M$-$H$ curves,
						are plotted on the panels (a) and (b), respectively.
						Also, the transition temperatures determined from
						dc susceptibility measurements (not shown) are plotted on both (a) and (b). 
						The color map of 
						$\mathrm{d}M/\mathrm{d}H_\mathrm{eff}$ (Fig. \ref{M-H} (b))
						normalized with the maximum value is also superimposed on the phase diagrams.
						The points 
						$\mathrm{X_{1,2}}$  and $\mathrm{Y}$ indicate the temperature where the type of the phase transition changes (see the text).
						In the panel (c), transition points from both the IF and DF processes inside dashed rectangles in the panel (a) and (b) are plotted  
						and the area of the GH loops is roughly colored.
						Note that the line colors and line types are different in this panel.}
		\label{pha_dia}
	\end{figure*}

	Open hysteresis loop between the phases III and IV indicates a strong nonequilibrium effect and the existence of metastable states in the magnetic field range of the phases III and IV.
	This point was investigated further by measuring another several minor loops inside the GH loop at 2 K 
	in the range of the magnetic field spanning the plateaus of the phases I, III and IV.
	
	Figure \ref{minor_loop_wh} (a) shows the results of
	two minor loops A and B measured in the different range of $H_\mathrm{ex}$,
	$8\ \mathrm{kOe} \leq H_\mathrm{ex} \leq 20\ \mathrm{kOe}$ and
	$8\ \mathrm{kOe} \leq H_\mathrm{ex} \leq 21\ \mathrm{kOe}$, respectively.
	Each measurement was started from the phase III plateau by cooling the sample in the ZFC condition followed by applying 
	$H_\mathrm{ex} = 15$ kOe.
	The loops A and B were measured twice consecutively.
	The ``whole loop'' representing the hysteresis loop at 2 K in 
	Fig. \ref{M-H} (a) is also plotted in the figure.
	First, the loops were measured from the phase III plateau to each maximum field and every loop is almost identical to the whole loop.
	With decreasing field from the maxima, the magnetizations keep the high values and exhibit broad peaks
	as well as the minor loop shown in Fig. \ref{small_minor} (b).
	These broad peaks are maximum-field-dependent and seen at around $H_\mathrm{eff} =  16 $ kOe for the loop A and at around $H_\mathrm{eff} = 16.5$ kOe for the loop B.
	Around the magnetic field region of the phase III ($11\ \mathrm{kOe} \leq H_\mathrm{eff} \leq 15\ \mathrm{kOe}$), the magnetization does not go back to that of the phase III plateau in either minor loop.
	Instead, plateau-like structures with gentle slopes appear.
	The magnetizations of these plateaus are different from those of the phase III,
	indicating that there are multiple metastable states inside the GH loop. 
	In the lower field range, the curves become steeper, merge with the whole loop and, eventually, reach the phase I plateau.
	With increasing the field again from the minimum value, both minor loops trace the whole loop with a small deviation.
	In the following 2nd cycles, the magnetizations in the DF branches tend to be smaller than those in the 1st cycles
	but their behaviors are almost the same as a whole.

	Furthermore, since the magnetizations in the minor loops depend on the number of the cycle, $n$, 
	it is indicated that there exist some higher-order nonequilibrium effects.
	To examine this effect in more detail, regarding the loop A, 
	the cycle was repeated 8 times.
	Fig. \ref{minor_loop_wh} (b) shows the difference in magnetization between the 
	$n$th and $(n+1)$th cycle and	
	the loop becomes unchanged at the 8th cycle, indicating the system reaches a steady state
	
	\ \\
	\noindent \textbf{$\bullet$ Phase diagram}

	Figure \ref{pha_dia} shows the
	$H_\mathrm{eff} \mathchar`- T$ phase diagrams of $\mathrm{DyRu_2Si_2}$ based on the hysteresis loop measurements (Fig. \ref{M-H}) and the temperature dependences of dc susceptibility.
	The results in the IF and DF branches are plotted separately in the panels (a) and (b), respectively.
	The phase diagram of the IF process is identical to the one reported in the earlier work 
	\cite{J.Phys.Cond.7.1889}. 
	The guide-to-eyes phase lines are also drawn.
	On the phase lines, the points $\mathrm{X_{1,2}}$ and $\mathrm{Y}$  indicate the temperatures where the type of the phase transition changes based on the results of magnetization curves (Fig. \ref{M-H}) and minor loops (Fig. \ref{small_minor}).
	Above $\mathrm{X_{1,2}}$, the phase lines are 
	second-order phase transitions and, below them, these lines
	become first-order ones with the appearance of the hysteresis.
	Regarding the phase lines of the phase IV, below 
	$\mathrm{Y}$, the phase lines cannot be classified into either conventional first- or second-order phase transition.
	The panel (c) is the enlarged figure of the area around the phase IV and III at low temperature, 
	which is indicated by the dashed rectangle in the panel (a) and (b), 
	and the transition points from both the IF and DF branches are plotted.
	In this panel, the area where the GH loops are observed is roughly colored.
	It is even clear that a part of the area of the phase III and whole of the phase IV doesn't exist in the DF process.
	It is indicated that any states in this colored area, 
	no matter those appearing either in the IF or DF processes, can be metastable.

	\subsection{Relaxations at the temperature and magnetic field of the phases III and IV}

	\begin{figure*}[t]
		\begin{tabular}{cc}
		\centering
			\begin{minipage}[t]{0.45\linewidth}
			\centering
			\includegraphics[height=5cm]{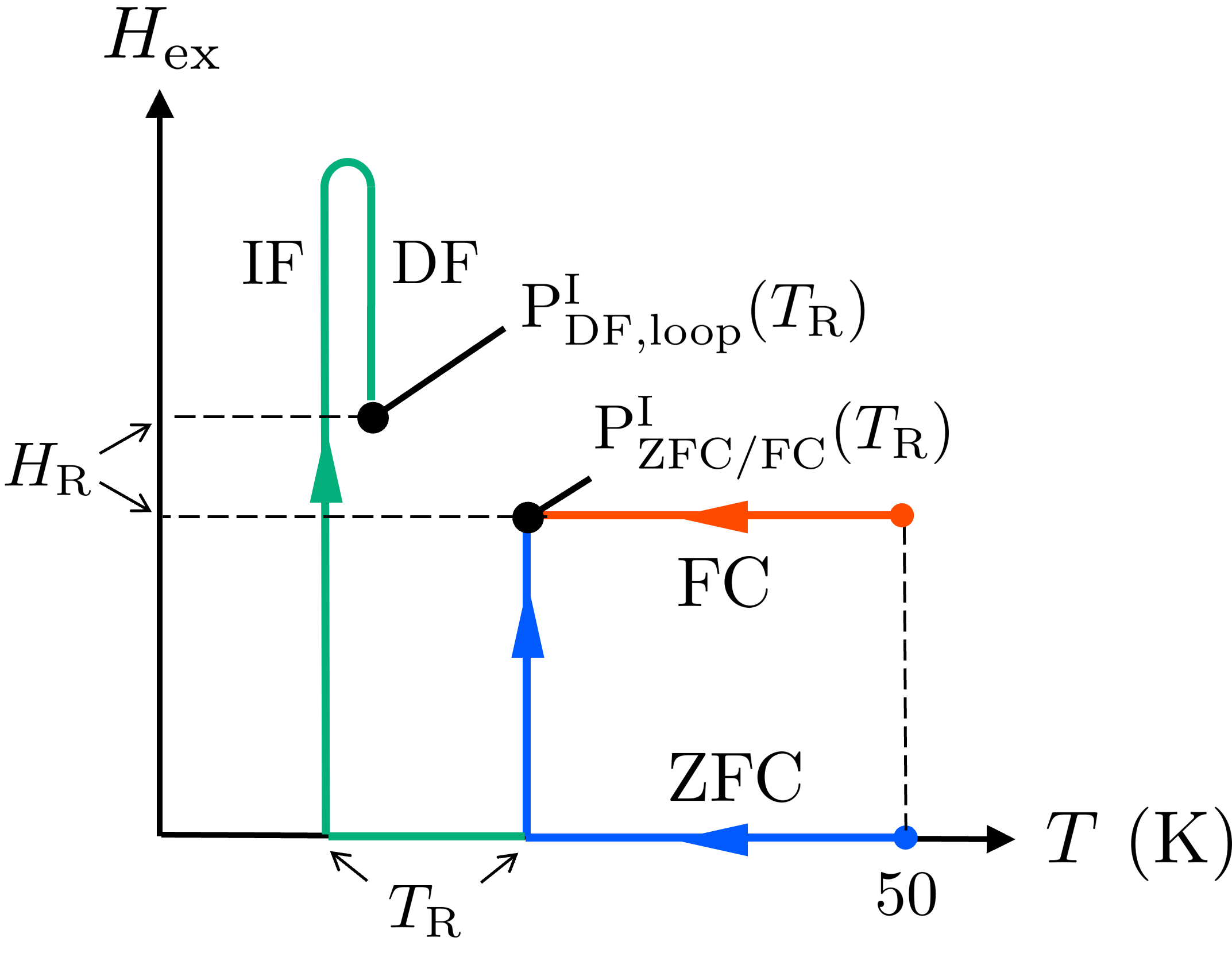}	\\ 			
			\centering {\small (a)}
			\end{minipage}

			\begin{minipage}[t]{0.45\linewidth}
			\centering
			\includegraphics[height=5cm]{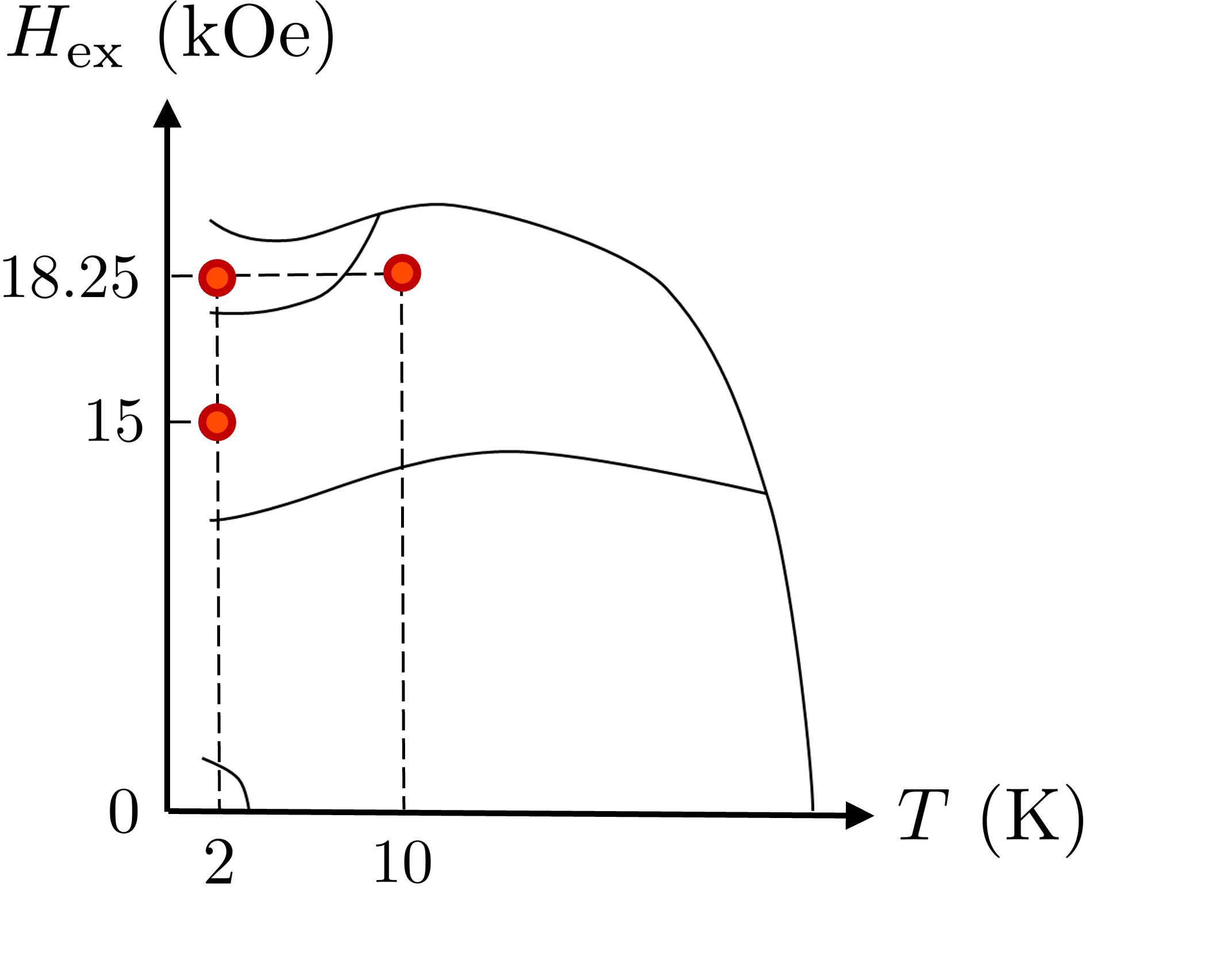}	\\ 
			\centering {\small (b)}
			\end{minipage}
		\end{tabular}

		\caption{(Color online) 
			(a) Schematized picture on the preparation of the initial states and their abbreviations in the $H_\mathrm{ex} \mathchar`- T$ diagram.
			(b) The positions of the initial states in the schematized phase diagram of the IF process.
				}
		\label{initial_states}
	\end{figure*}

	In this subsection, to determine whether the states in the colored area in Fig. \ref{pha_dia} (c)
	are metastable or not,
	we investigated whether these states, prepared in various ways, exhibit magnetic relaxation or not.  
	Here, we first explain how the initial states of relaxations were prepared.
	In Fig. \ref{initial_states} (a), the preparation of the initial states in the
	$H_\mathrm{ex} \mathchar`- T$ diagram and their abbreviations, 
	$\mathrm{P^I_{ZFC/FC}}(T_\mathrm{R})$ and
	$\mathrm{P^I_{DF, loop}}(T_\mathrm{R})$, are schematized.
	The initial states at temperature $T_\mathrm{R}$ and external field $H_\mathrm{R}$ were prepared in three kinds of process as schematized in the figure.
	Also, in Fig. \ref{initial_states} (b), the positions of those initial states are shown in the schematized $H_\mathrm{ex} \mathchar`- T$ phase diagram of the IF process.
	The rules for the abbreviations are as follows.
	The capital ``P" refers to the name of the phase that appears in the IF process at particular 
	$T_\mathrm{R}$ and $H_\mathrm{R}$,
	 which are ``III" or ``IV" in this study.
	The superscript of ``I" stands for ``initial states".
	The subscripts indicate the processes of the preparation of the initial states as follows.
	The state
	$\mathrm{P^I_{ZFC}}(T_\mathrm{R})$ is prepared by cooling 
	the sample in the ZFC condition down to $T_\mathrm{R}$ followed by applying 
	$H_\mathrm{ex} = H_\mathrm{R}$.
	The state
	$\mathrm{P^I_{FC}}(T_\mathrm{R})$ is prepared by cooling 
	the sample in the FC condition down to $T_\mathrm{R}$ under the field of 
	$H_\mathrm{ex} = H_\mathrm{R}$.
	Note that these states, $\mathrm{P^I_{ZFC/FC}} (T_\mathrm{R})$,
	are those in the plateau of the phase III or IV. 
	The states
	$\mathrm{P^I_{DF,loop}} (2\ \mathrm{K})$ are the states in the DF process of a particular ``loop", which are the whole loop, the minor loop A, or C.
	In the case of the whole loop, the abbreviation is
	$\mathrm{P^I_{DF,W}} (2\ \mathrm{K})$.
	This state is prepared by cooling the sample in the ZFC condition down to $T_\mathrm{R}$, applying the field up to 
	$H_\mathrm{ex}=30\ \mathrm{kOe}$, and decreasing the field down to 
	$H_\mathrm{ex} = H_\mathrm{R}$ at a particular time rate.
	In the case of the loop A, the abbreviation is
	$\mathrm{P^I_{DF,A}} (2\ \mathrm{K})$.
	This state is prepared by measuring the loop A in the same procedure as described so far.
	For the loop C, 
	which was measured in the same way as the loop A but with the lower maximum field of 
	$H_\mathrm{ex} = 19.25$ kOe (see Fig.
	\ref{IV_relax} (b)),
	the abbreviation is
	$\mathrm{IV^I_{DF, C}(2\ K)}$.
	Note that each relaxation measurement was started just after reaching each initial state.
	The initial states of the  relaxations in this study are listed in Table I.

	\begin{table}[t]
		\centering
 		\tblcaption{
 				Initial states of relaxation.
				}
		\label{states_table}
		\renewcommand{\arraystretch}{1.4}
		\begin{tabular}{p{30mm}p{20mm}} \hline
			\multicolumn{1}{c}{Initial states} & \multicolumn{1}{c}{$H_\mathrm{R}$ (kOe)}\\ 
			\hline
			$\mathrm{III^I_{ZFC} (2\ K)}$ & 15 \\
			$\mathrm{III^I_{ZFC/FC} (10\ K)}$ & 18.25 \\
			$\mathrm{III^I_{DF,A} (2\ K)}$ & 15 \\
			$\mathrm{III^I_{DF,W} (2\ K)}$ & 15 \\
			$\mathrm{IV^I_{ZFC/FC}} (T_\mathrm{R})$ & 18.25 \\
			$\mathrm{IV^I_{DF,A} (2\ K)}$ & 18.25 \\
			$\mathrm{IV^I_{DF,C} (2\ K)}$ & 18.25 \\
			$\mathrm{IV^I_{DF,W} (2\ K)}$ & 18.25 \\
			 \hline
		\end{tabular}
	\end{table}

	\begin{figure*}[t]
		\centering
		\begin{tabular}{cc}
			\begin{minipage}[c]{0.45\linewidth}
				\centering
				\includegraphics[height=5.5cm]{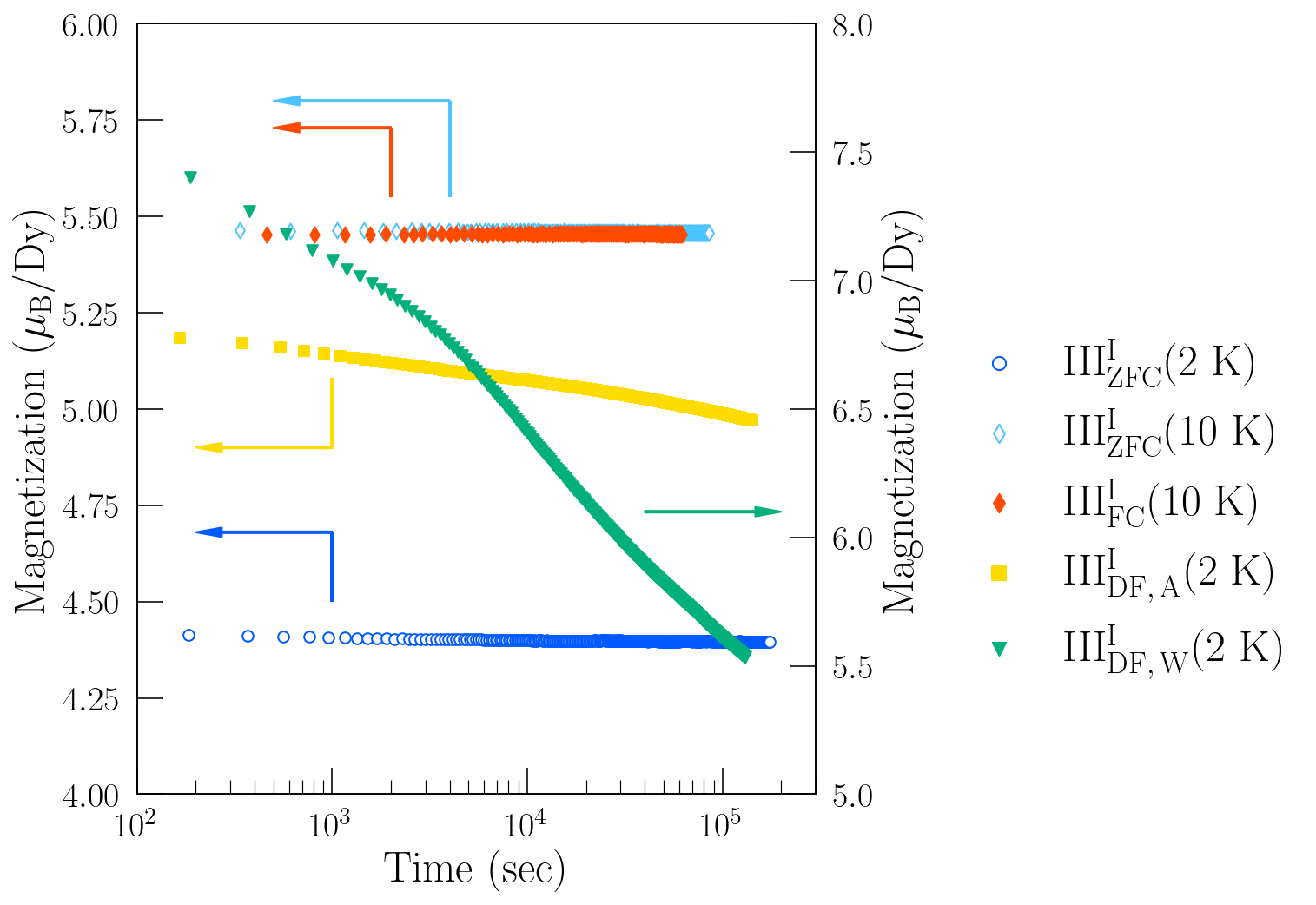}
				\\ 
				\centering {\small (a) }
			\end{minipage}
			\hspace{10pt}
			\begin{minipage}[c]{0.4\linewidth}
				\centering
				\includegraphics[height=5.5cm]{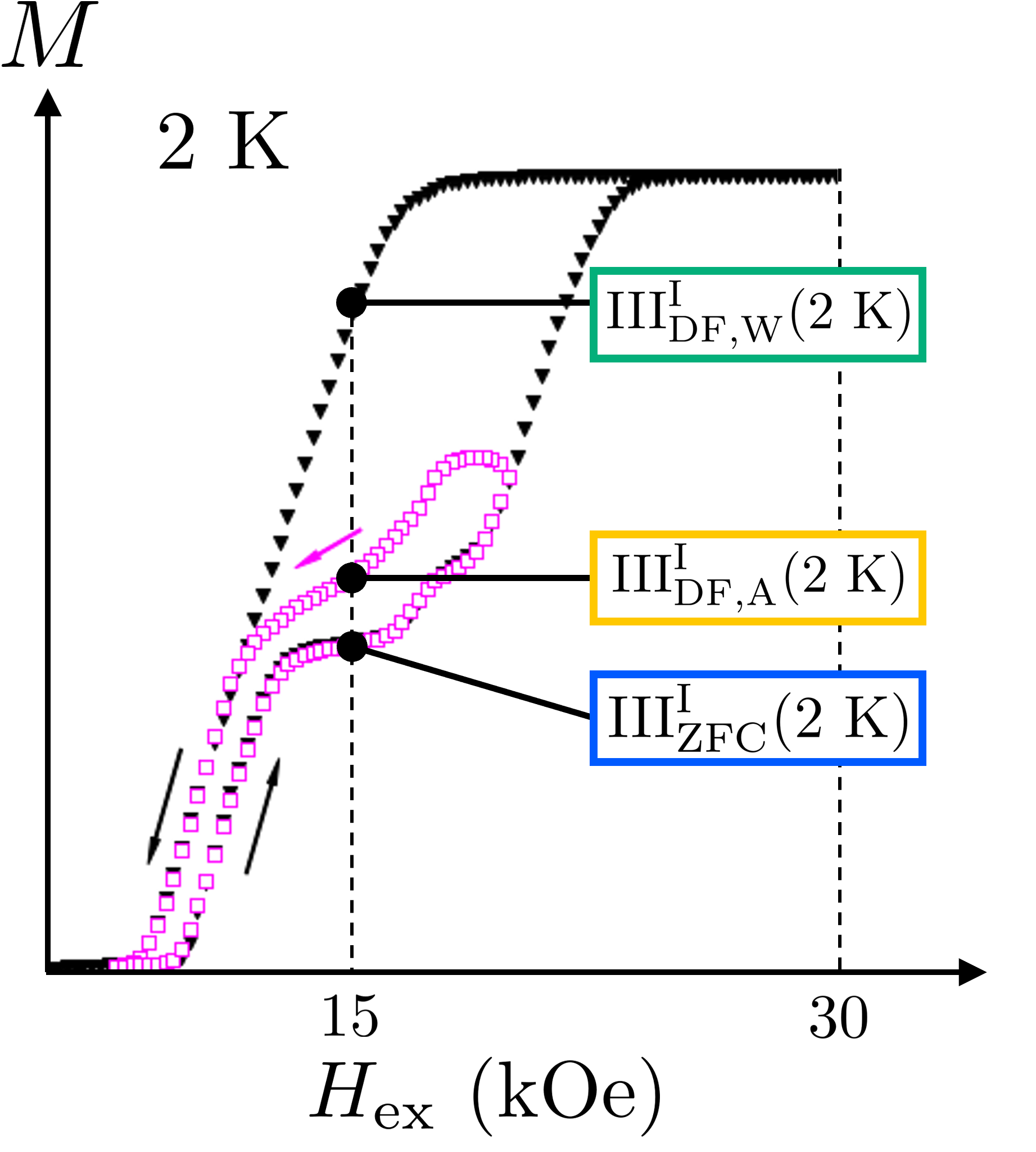}
				\\ 
				\centering {\small (b) }
			\end{minipage}
		\end{tabular}
		\caption{(Color online) 
					(a) Relaxation from each initial state at $H_\mathrm{ex} = 15\ \mathrm{kOe}$.
					(b) Positions of the initial states on the hysteresis loops at 2 K.
					Note that the external magnetic field, not the effective field, is used here.
				}
		\label{III_relax}
	\end{figure*}

		\begin{figure*}[t]
			\centering
			\begin{tabular}{ccc}
				\begin{minipage}[c]{0.45\linewidth}
					\centering
					\includegraphics[height=5.5cm]{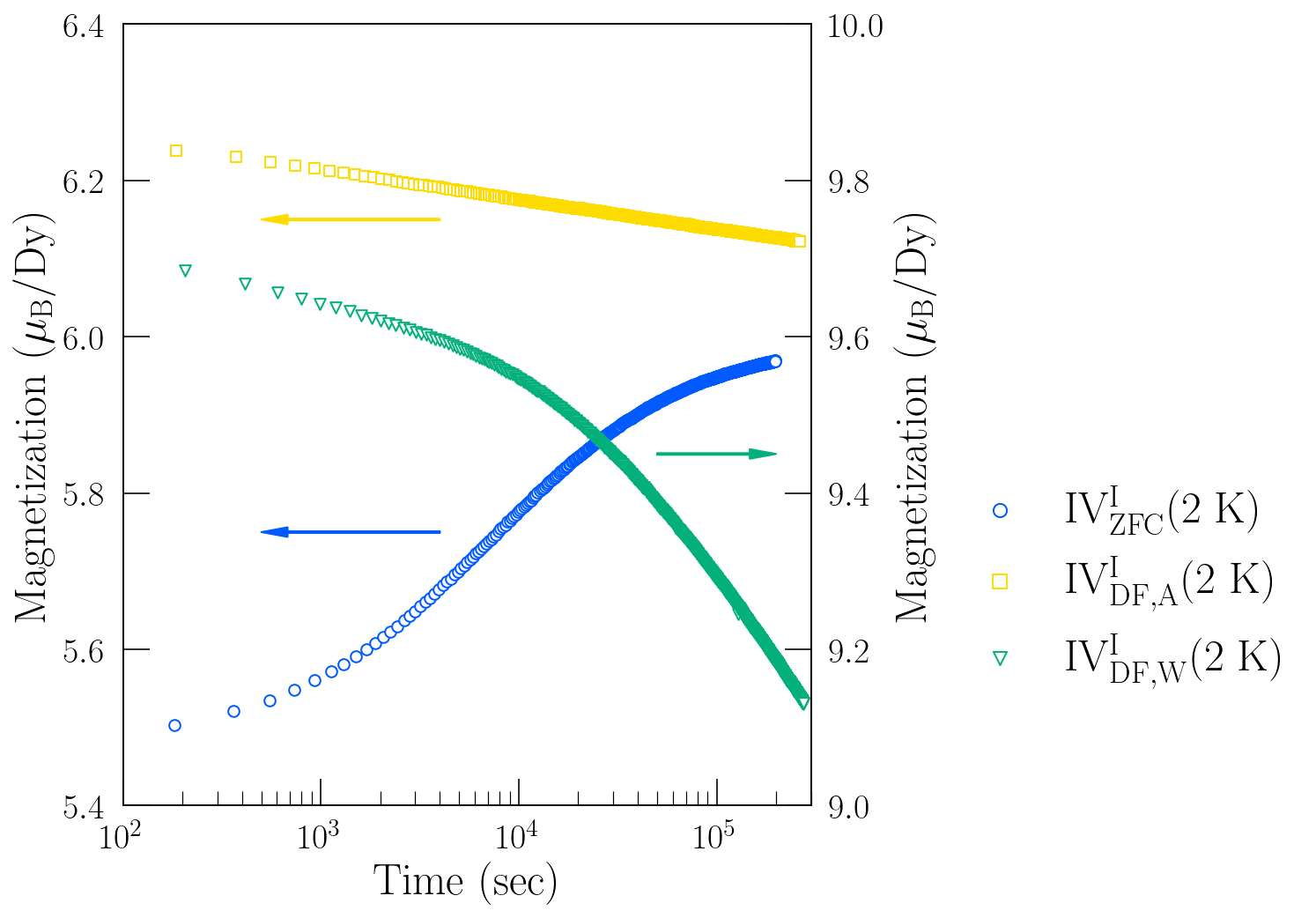}
					\\ 
					\centering {\small (a) }
				\end{minipage}
				\hspace{10pt}
				\begin{minipage}[c]{0.4\linewidth}
					\centering
					\includegraphics[height=5.5cm]{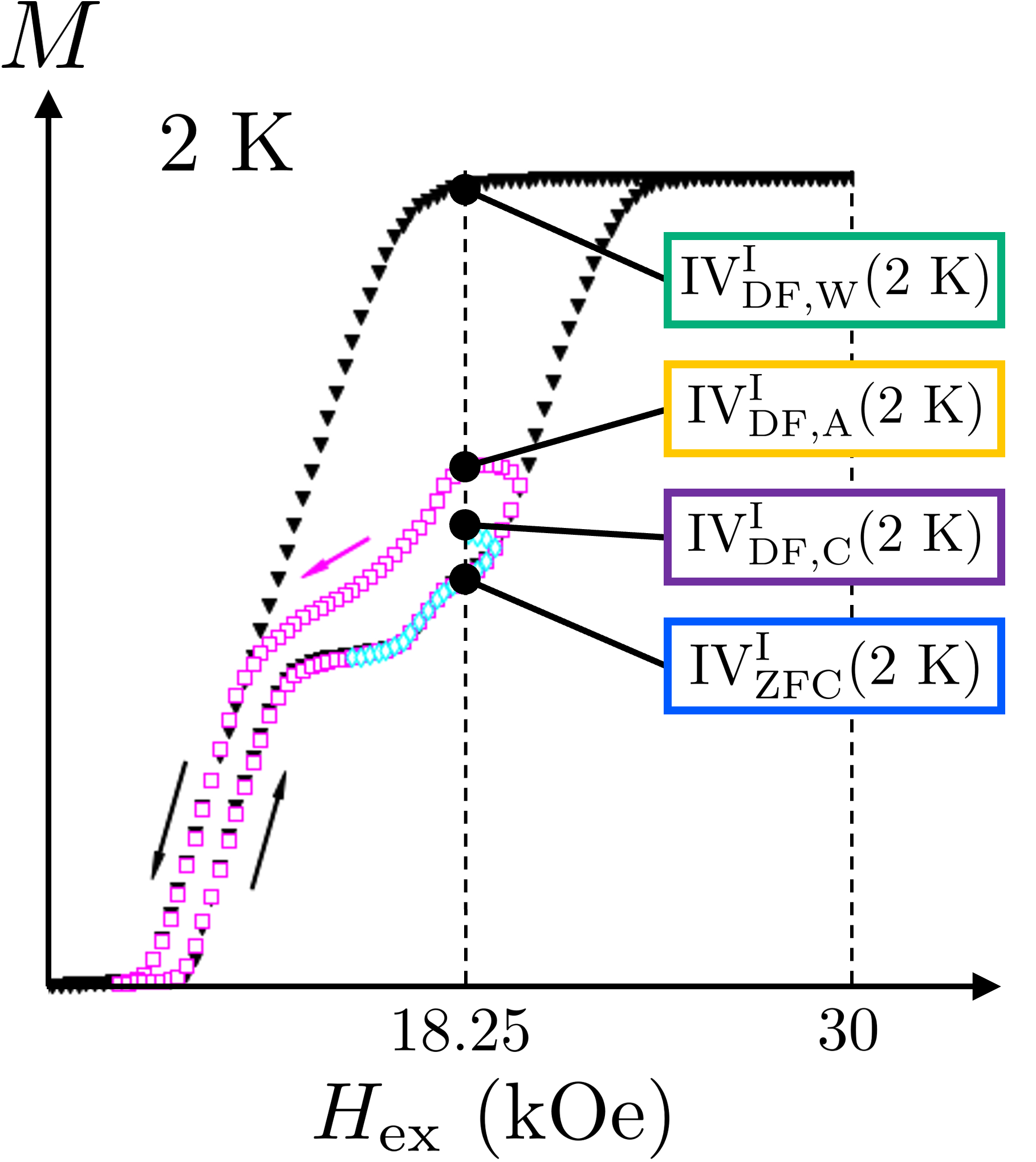}
					\\ 
					\centering {\small (b) }
				\end{minipage}
			\end{tabular}
			\caption{(Color online) 
					(a) Relaxation from each initial state at $H_\mathrm{ex}=18.25$ kOe.
					(b) Positions of the initial states on the hysteresis loops at 2 K. 
					Note that the external magnetic field, not the effective field, is used here.
						}
			\label{IV_relax}
		\end{figure*}

		Figure \ref{III_relax} (a) shows the magnetization relaxations from several initial states 
		of $\mathrm{P=III}$		
		(see Table \ref{states_table} and Fig. \ref{III_relax} (b)).
		No remarkable relaxations are observed from the initial states 
		$\mathrm{III^I_{ZFC}(2\ K)}$ and $\mathrm{III^I_{ZFC/FC}(10\ K)}$, indicating that 
		these plateau states in the IF branches are the thermal equilibrium states of the phase III.
		On the other hand, long time relaxations are observed from the initial states
		$\mathrm{III^I_{DF,W}(2\ K)}$ and
		$\mathrm{III^I_{DF, A}(2\ K)}$, which indicates that 
		these states in the DF branches are metastable rather than thermal equilibrium.
		In these relaxations, the magnetizations keep decreasing for over $10^5$ sec and do not asymptotically approach any value within the measurement time.

		Figure \ref{IV_relax} (a) shows the magnetization relaxations from several initial states 
		of $\mathrm{P=IV}$
		(see Table \ref{states_table} and Fig. \ref{IV_relax} (b)).
		A long-time relaxation is observed from the initial state on the phase IV plateau,
		$\mathrm{IV^I_{ZFC}(2\ K)}$, and the magnetization increases with time.
		This is the most striking difference compared to the relaxations in the phase III plateaus,
		from $\mathrm{III^I_{ZFC}(2\ K)}$ and $\mathrm{III^I_{ZFC/FC}(10\ K)}$.
		Therefore, it is considered that the phase IV plateau in the IF branch is metastable, not thermal equilibrium.
		Long-time relaxations are also observed from the initial states of
		$\mathrm{IV^I_{DF, W}(2\ K)}$ and $\mathrm{IV^I_{DF,A}(2\ K)}$, which indicates that 
		these states in the DF branches are metastable
		as well as the initial states		
		$\mathrm{III^I_{DF, W}(2\ K)}$ and
		$\mathrm{III^I_{DF, A}(2\ K)}$.
		In these relaxations in the phase IV, the magnetizations do not saturate within the measurement time. 
		We note that the increase in magnetization from the initial state 
		$\mathrm{IV^I_{ZFC}(2\ K)}$ indicates that the broad peaks in the DF branches of the minor loops in 
		Fig. \ref{small_minor} (b) and \ref{minor_loop_wh} (a)
		are due to the remaining relaxation effect of the phase IV plateau or some nonequilibrium effects coming from the field increase
		in the IF branches.

		\subsection {Characterization of the nonequilibrium phenomena in the phase IV plateau}
		
		As described in the previous subsection, the phase IV plateau in the IF process of the GH loop is metastable in contrast to the phase III plateaus.
		In this subsection, the nonequilibrium phenomena of the phase IV are investigated further through relaxation measurements 
		from several different $T_\mathrm{R}$. 
		Also, the magnetization of the true equilibrium state is estimated by extrapolating the relaxations to time infinity.
		Furthermore, the temperature dependences of the susceptibility from the final states of the relaxations are measured to
		elucidate whether the long-time relaxation is the intrinsic nature of the phase IV or not.

		\ \\
		\noindent {\textbf{$\bullet$ Relaxations at different $T_\mathrm{R}$}}
	
			\begin{figure*}[t]
				\centering
				\begin{tabular}{cccc}
					\begin{minipage}[b]{0.45\linewidth}
						\centering
						\includegraphics[width=5cm]{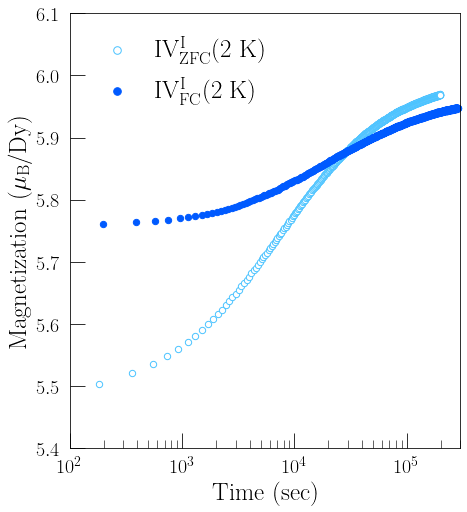}
						\\ 
						\centering {\small (a) }
					\end{minipage}
					\begin{minipage}[b]{0.45\linewidth}
						\centering
						\includegraphics[width=5cm]{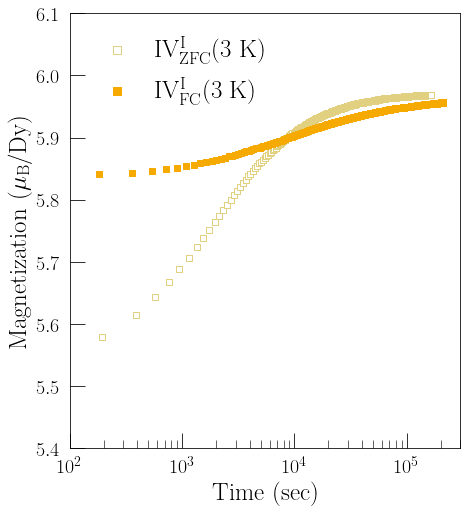}
						\\ 
						\centering {\small (b) }
					\end{minipage}
					\\
					\begin{minipage}[c]{0.3\linewidth}
						\centering
						\includegraphics[width=5cm]{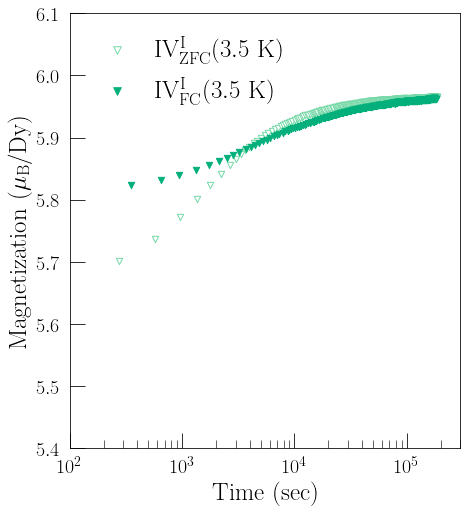}
						\\ 
						\centering {\small (c) }
					\end{minipage}
					\hspace{25pt}
					\begin{minipage}[c]{0.3\linewidth}
						\centering
						\includegraphics[width=5cm]{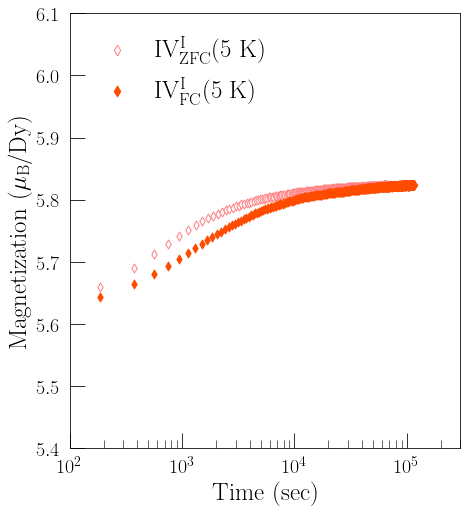}
						\\ 
						\centering {\small (d) }
					\end{minipage}
					\hspace{25pt}
					\begin{minipage}[c]{0.25\linewidth}
						\centering
						\includegraphics[width=4.5cm]{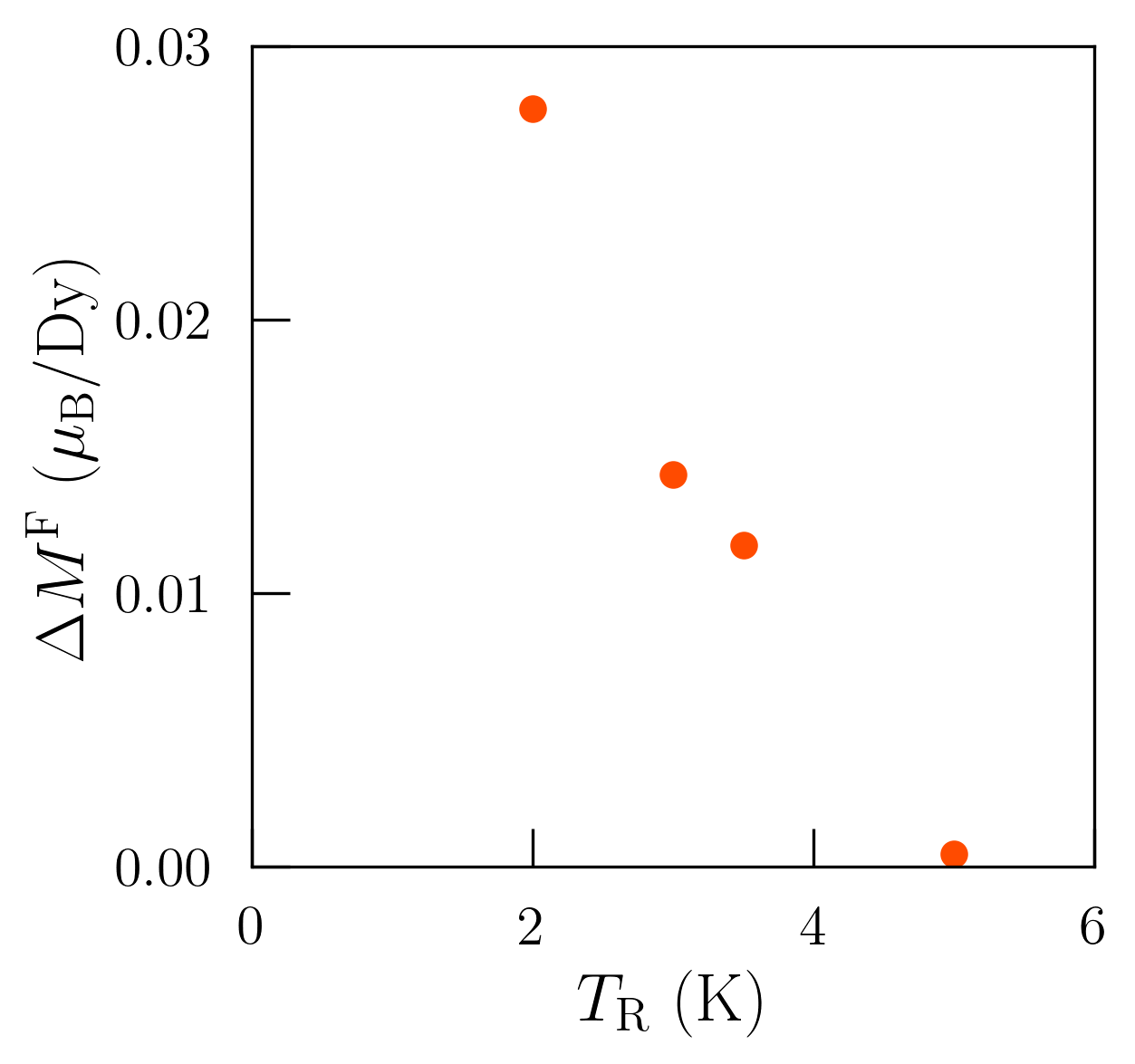}
						\\
						\centering {\small (e) }
					\end{minipage}
				\end{tabular}
				\caption{ (Color online) 
							Magnetization relaxations from the initial states
							$\mathrm{IV^I_{ZFC/FC}} (T_\mathrm{R})$
							for (a) $T_\mathrm{R}=2.0 \mathrm{K}$, (b) 3.0 K, (c) 3.5 K and (d) 5.0 K.
							(e) The temperature dependence of the difference between the ZFC and FC magnetizations at the final states, 
							$\Delta M^\mathrm{F}$.
							}
				\label{relax_IV}
			\end{figure*}

			\begin{figure*}[t]
				\centering
				\begin{tabular}{cc}
					\begin{minipage}{0.4\linewidth}
						\centering
						\includegraphics[width=8cm]{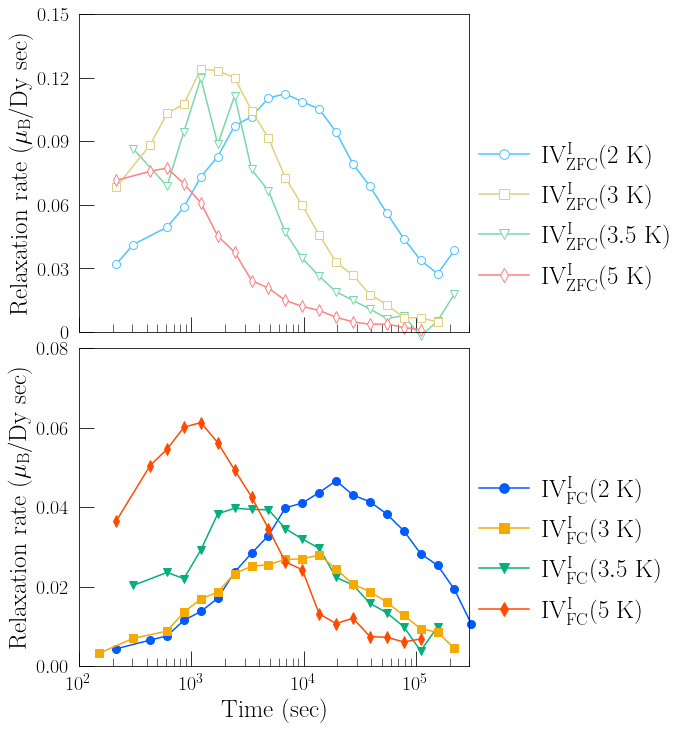}
						\\
						\centering {\small (a)}
					\end{minipage}
					\begin{minipage}{0.4\linewidth}
						\centering
						\includegraphics[width=5cm]{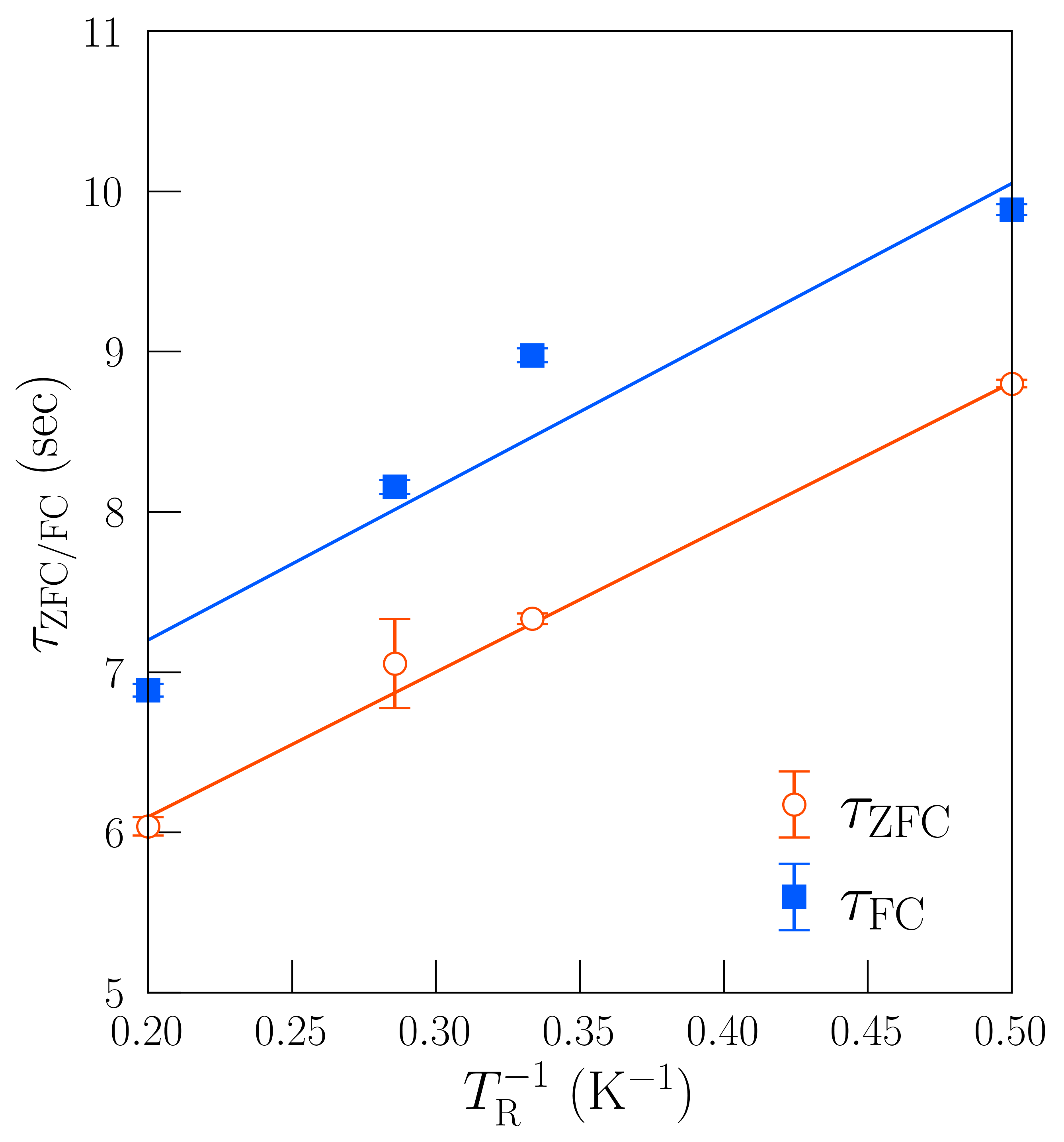}
						\\
						\centering {\small (b)}
					\end{minipage}
				\end{tabular}
				\caption{(Color online) 
								(a) Relaxation rates calculated from the magnetization relaxations shown in Fig. \ref{relax_IV}.
								The upper and lower panels show the results of the ZFC and FC conditions.
								(b) Arrhenius plot of the estimated characteristic relaxation times  of the relaxations against $T_\mathrm{R}$,
								$\mathrm{ln} \tau_\mathrm{ZFC/FC}$ vs $T_\mathrm{R}^{-1}$.
								The solid lines represent the Arrhenius law, given by eq. (3). 
								}
				\label{relax_rate}
			\end{figure*}

		The relaxations in the phase IV from the initial states
		$\mathrm{IV^I_{ZFC}}(T_\mathrm{R})$ and
		$\mathrm{IV^I_{FC}}(T_\mathrm{R})$ for
		$T_\mathrm{R} = 2.0, 3.0, 3.5$ K and $5.0 \ \mathrm{K}$
		were measured (Figs. \ref{relax_IV} (a)-(d)).
		Hereafter, $T_\mathrm{R}$ is called ``relaxation temperature'' and the relaxations from the initial states
		$\mathrm{IV^I_{ZFC/FC}}(T_\mathrm{R})$ are called ``the ZFC/FC relaxation at $T_\mathrm{R}$''.
		From every initial state, even those prepared in the FC condition, 
		magnetizations increase monotonically with time for 
		longer than $10^5 \ \mathrm{sec}$,
		which do not saturate within the measurement time.
		This indicates that these initial states are 
		all metastable and the true equilibrium state exists elsewhere. 
		Surprisingly, comparing each pair of the ZFC and FC relaxations at the same $T_\mathrm{R}$, 
		the magnitude relationship of the magnetization  
		is reversed in the middle of the relaxations, where the magnetization of the FC relaxation is larger in the initial states but 
		that of the ZFC relaxation is larger in the final states of the measurement. 
		This is a highly nontrivial and counterintuitive outcome, since it is usually expected that 
		the FC magnetization is always larger than the ZFC magnetization.
		This behavior resembles the Mpemba effect
		\cite{PhysEdu.4.1969}, where the initially hotter system starts to freeze faster than the colder system, and this analogy is extensively discussed in Sect. 4.
		Also, the difference between the magnetizations of ZFC and FC relaxations at $t \approx 10^5$ sec, which is almost the final states of the relaxation, 
		$\Delta M \mathrm{^{F}}$, is plotted against $T_\mathrm{R}$ in Fig. \ref{relax_IV} (e).
		It is finite at $T_\mathrm{R} = 2,\ 3,\ 3.5\ \mathrm{K}$ and almost zero at 5 K.
		$\Delta M^\mathrm{F}$ increases with decreasing the temperature.
		Even though the measurements are done in the finite time, the difference in magnetizations at the final states of relaxations implies that 
		there are multiple equilibrium states in the phase IV.
		The intersecting relaxations and the finite $\Delta M \mathrm{^{F}}$ indicate that the different cooling protocols lead to completely different relaxation processes and final states.

			The relaxation rate
			$S(t)$ derived from the magnetization relaxation (Figs. \ref{relax_IV} (a)-(d)) is plotted in Fig. \ref{relax_rate} (a).
			$S(t)$ is defined as
			\begin{equation}
				S(t) := 
					\frac
					{d M}
					{d \mathrm{ln}t} \, .
			\end{equation}
			
			\noindent
			The figure shows the average values of $S(t)$ for every constant interval of $t$ in the logarithmic scale.
			As seen in Fig. \ref{relax_rate} (a), every $S(t)$ exhibits a broad peak structure.
			The relaxation rate is proportional to the distribution of relaxation time, so its peak time 
			represents the characteristic relaxation time of the relaxation
			\cite{PhysRevLett.51.911}.
			The characteristic time $\tau$ was estimated by fitting the curves in the vicinity of each peak with the function 
			$y = \mathrm{a}(\mathrm{ln}t - \mathrm{ln}\tau)^2 + \mathrm{b}$, where $\mathrm{a, b\ and\ \tau}$ are the fitting parameters.
			In Fig. \ref{relax_rate} (b),
			the estimated characteristic times of the ZFC and FC relaxations, $\tau_\mathrm{ZFC/FC}$, 
			are plotted against $T_\mathrm{R}$.
			The temperature dependences of 
			$\tau_\mathrm{ZFC/FC}$ are approximately described by the Arrhenius low;

			\begin{equation} 
			\tau_\mathrm{ZFC/FC}=\tau_{0, \mathrm{ZFC/FC}}\, \mathrm{exp}
			\Bigg( \frac{E_\mathrm{ZFC/FC}}{T} \Bigg) ,
			\end{equation}

			\noindent
			indicating that the relaxation is thermally activated, as usually expected for classical spin systems.
			As the fitting parameters by eq. (3), the activation energies and pre-exponential factors
			are determined as follows, 	
			$E_\mathrm{ZFC}=9.03 \pm 0.17\ \mathrm{K}$,
			$E_\mathrm{FC}=9.49 \pm 0.16\ \mathrm{K}$,
			$\tau_\mathrm{0, ZFC}=73.0 \pm 1.0\ \mathrm{sec}$ and
			$\tau_\mathrm{0, FC}=200.5 \pm 1.0 \ \mathrm{sec}$.
			Comparing the ZFC and FC relaxations at the same $T_\mathrm{R}$,
			the peak time of the relaxation rate is always shorter for the ZFC condition,
			and moreover, the entire ZFC relaxation rate is biased towards a shorter time region.
			This indicates that the ZFC relaxation is faster than the FC one, which leads to the overtaking of the FC relaxation by the ZFC one.

			\ \\
			\noindent \textbf{$\bullet$ The thermal equilibrium state of the phase IV and the estimation of its magnetization}

			In the phase IV, since the relaxation is a very long-time, the equilibrium magnetization is not able to be observed in the experimental timescale.
			Related to this, it is difficult to determine whether the number of equilibrium states is
			unique or multiple and both are possible.
			As discussed above, finite 
			$\Delta M ^\mathrm{F}$ at 
			$T_\mathrm{R}=2,\ 3,\ 3.5\ \mathrm{K}$ suggest that multiple thermal equilibrium states 
			with different magnetization exist.
			On the other hand, 
			one can expect a unique equilibrium state as follows.
			$\Delta M ^\mathrm{F}$ decreases with increasing temperature and 
			becomes almost zero at 5 K, and the magnetization
			relaxation is thermally activated.
			Considering this, it can be expected that 
			after a sufficiently long time,
			$\Delta M^\mathrm{F}$ would asymptotically approach zero and
			the magnetization would saturate to the same value from any initial state.
			Therefore, in this sense, the thermal equilibrium state is thought to be unique.

\begin{figure*}[t]
		\centering
		\begin{tabular}{cc}
			\begin{minipage}{0.45\linewidth}
				\centering
				\includegraphics[height=6cm]{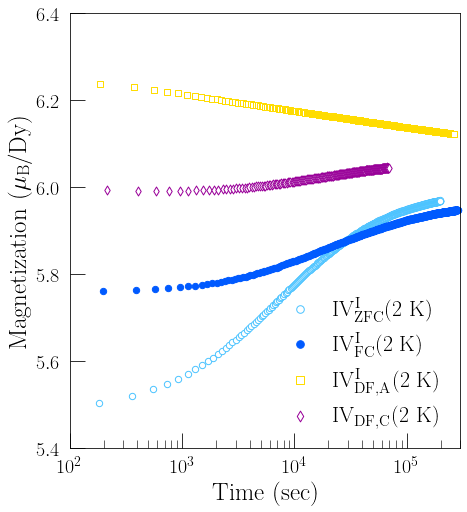}
				\\
				\centering {\small (a)}
			\end{minipage}
			\begin{minipage}{0.45\linewidth}
				\centering
				\includegraphics[height=6cm]{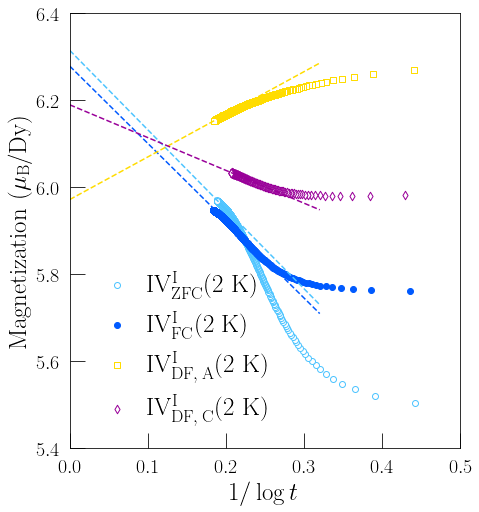}
				\\
				\centering {\small (b)}
			\end{minipage}
		\end{tabular}
 								\caption{(Color online) 
						(a)
						Relaxations from each initial state at 2 K in the phase IV.
						(b)
						Relaxations plotted against								
						$1/\mathrm{ln}t$.
						The dashed lines represent the linear extrapolations to 
						$\mathrm{ln}t \rightarrow 0$, namely, to time infinity.
						}
		\label{determination_of_eqstate}

	\end{figure*}

			To verify these two scenarios whether the number of the equilibrium state is unique or multiple, 
			the equilibrium magnetization of the phase IV is estimated by extrapolating some of the relaxations at 2 K to the time infinity.
			Fig. 
			\ref{determination_of_eqstate} (a) shows the relaxations from the initial states
			$\mathrm{IV^I_{ZFC/FC}(2\ \mathrm{K})}$, $\mathrm{IV^I_{DF, A}}(2\ \mathrm{K})$ and  $\mathrm{IV^I_{DF,C}(2\ K)}$. 
			The first three relaxations have already been described (Figs. \ref{IV_relax} (a) and \ref{relax_IV} (a)).
			From $\mathrm{IV^I_{DF,C}(2\ K)}$, the magnetization increases monotonically.
			This relaxation is shown here for the reference for the extrapolation 
			because the magnetization is intermediate compared to the other three relaxations and the change of the magnetization is smaller than those of the other relaxations.
			For the estimation of the infinite-time limits of these magnetizations, 
			the relaxations are plotted against 
			$1/ \mathrm{ln} t$ in Fig. \ref{determination_of_eqstate} (b)
			and the linear extrapolations to 
			$1/\mathrm{ln}t \rightarrow 0$, namely, to time infinity, was done as indicated with the dotted lines.
			The $1/ \mathrm{ln}t$ extrapolation is a convenient method attempted in spin glass relaxations
			\cite{Mamiya.Phil.Mag.Lett.2005.85}.
			The extrapolated values of magnetization are summarized in Table \ref{table_extrap}.
			The relaxations from the initial states
			$\mathrm{IV^I_{ZFC/FC}(2\ \mathrm{K})}$ and $\mathrm{IV^I_{DF, C}(2\ K)}$ indicate that the equilibrium magnetizations are 
			between 6.18 and 6.31 $\mu_\mathrm{B}$/Dy.
			On the other hand, the relaxation from the initial state  
			$\mathrm{IV^I_{DF, A}(2\ K)}$, whose magnetization is close to those above-mentioned extrapolated values,
			is extrapolated to the smaller value of 5.97 $\mu_\mathrm{B}$/Dy.
			The equilibrium value of magnetization could not be determined so accurately even with this extrapolation due to the limitation of the measurement time.
			At best, this result implies the scenario that multiple thermal equilibrium states with different magnetizations exist in the phase IV.

	\begin{table*}
		\centering
 					\tblcaption{
 									Infinit-time extrapolations of magnetizations measured from several initial states in the phase IV.
						}
		\label{table_extrap}
		\renewcommand{\arraystretch}{1.4}
		\begin{tabular}{ccccc} \hline
			initial states
			& $\mathrm{IV^I_{ZFC}(2\ K)}$
			& $\mathrm{IV^I_{FC}(2\ K)}$
			& $\mathrm{IV^I_{DF, A}(2\ K)}$
			& $\mathrm{IV^I_{DF, C}(2\ K)}$ \\ \hline
			extrapolated values
			($\mu\mathrm{_B/Dy}$)
			&6.31
			&6.27
			&5.97
			&6.18\\ \hline
		\end{tabular}
 	\end{table*}

				\begin{figure*}[t]
					\begin{tabular}{cc}
						\begin{minipage}{0.45\linewidth}
							\centering
							\includegraphics[height=6cm]{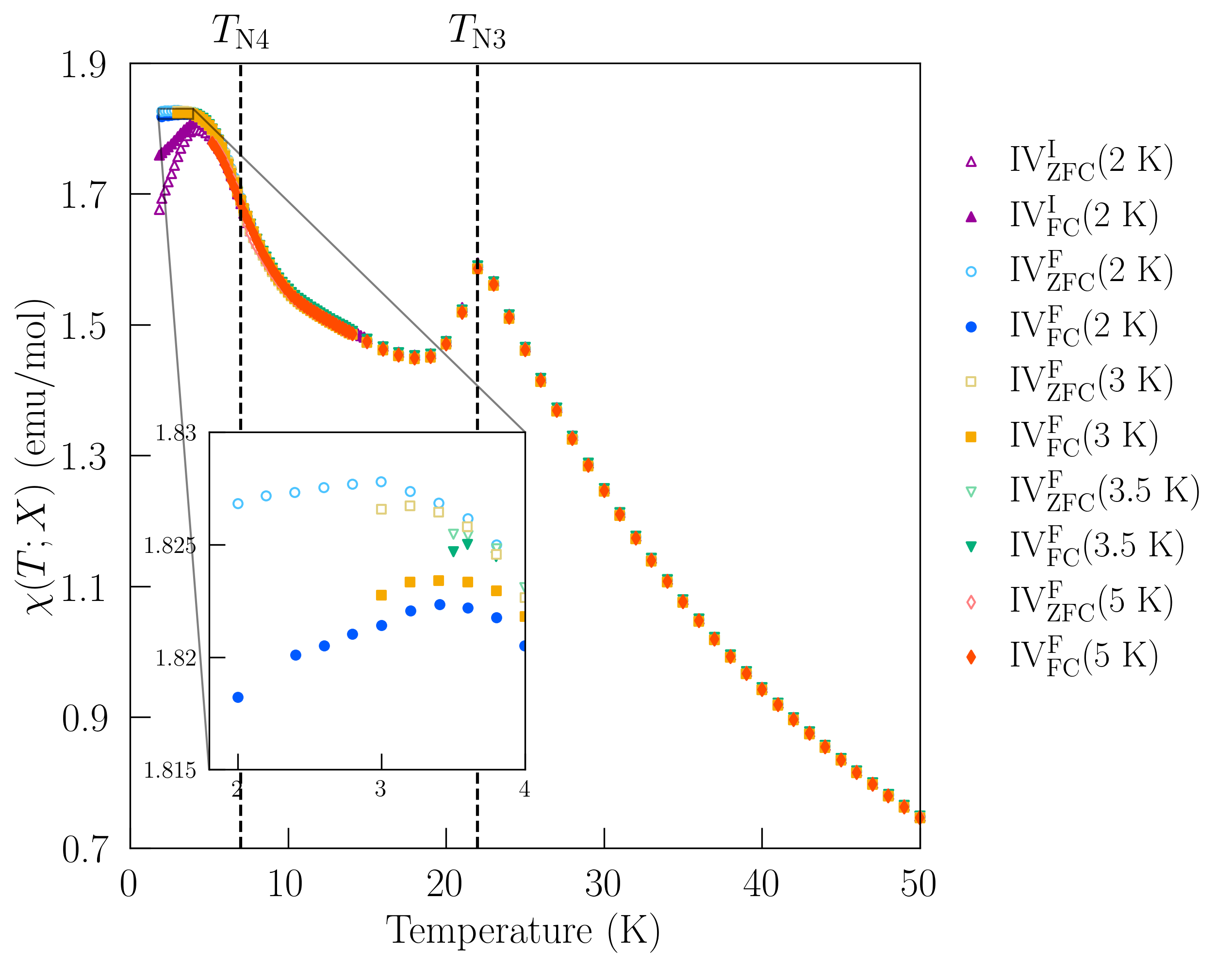}
							\\
							\centering {\small (a)}
						\end{minipage}
						\begin{minipage}{0.45\linewidth}
							\centering
							\includegraphics[height=6cm]{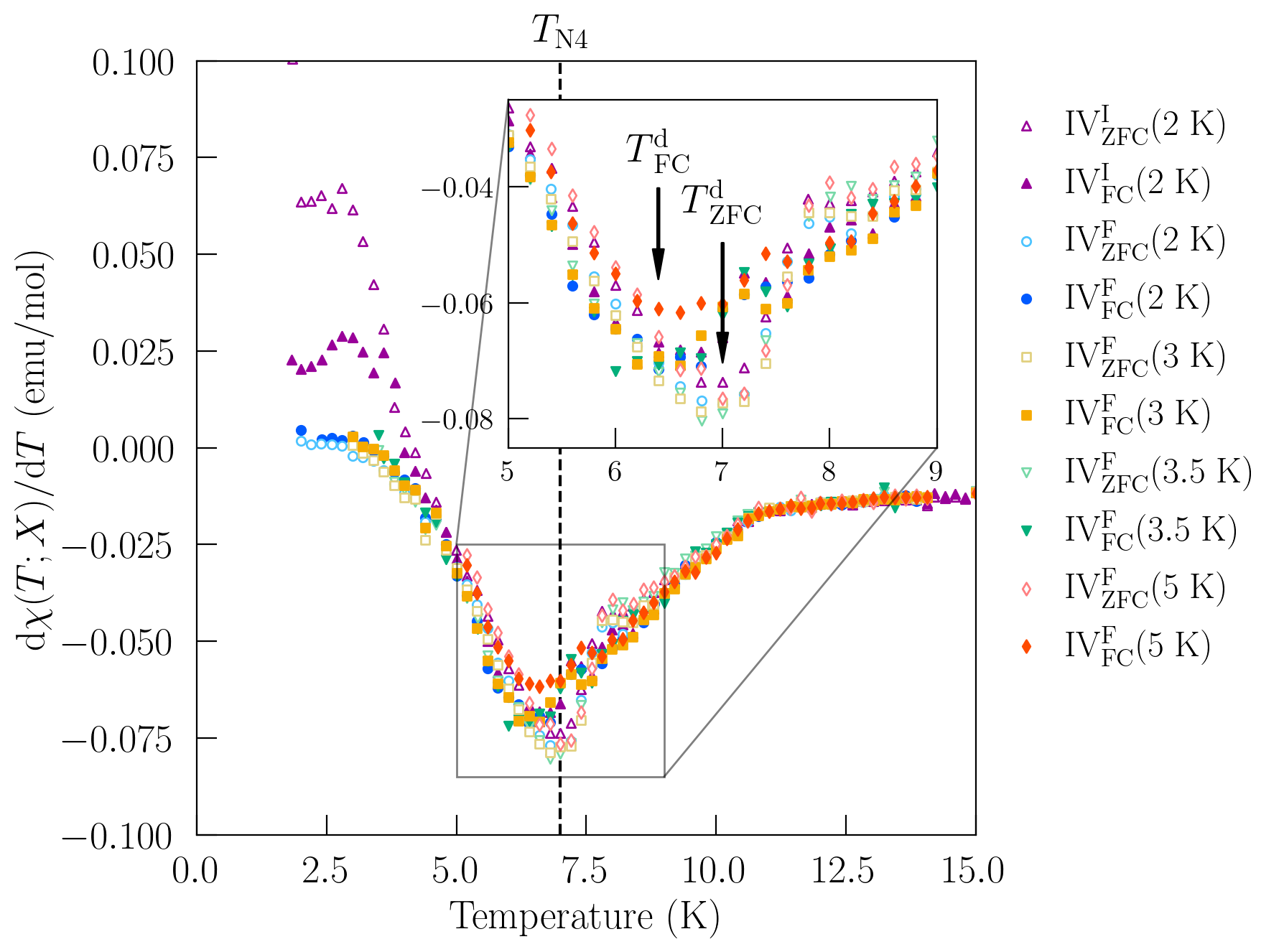}
							\\
							\centering {\small (b)}
						\end{minipage}
					\end{tabular}
					\caption{(Color online) 
						(a) Temperature dependences of the susceptibilities measured from the initial and final states of the relaxations.
						The inset is an enlarged figure of the low-temperature region. 
						(b) Temperature derivatives of the susceptibilities. 
						The inset is an enlarged figure in the vicinity of the dip temperatures.	
							}
				\label{chiF}
				\end{figure*}

\ \\

		\noindent \textbf{$\bullet$ Temperature dependences of susceptibility after relaxations in the phase IV}

			Even though extraordinarily long-time relaxation is observed in the phase IV, it is unclear whether this nonequilibrium phenomenon is the intrinsic feature occurring in the
			phase IV. 
			One possible and most likely interpretation for such a long-time relaxation is that the
			III-IV phase transition is the first-order phase transition,
			and, thus, the observed relaxations are those from the
			metastable phase III to the stable phase IV. 
			Here  this scenario is tested by measuring the temperature
			dependences of the susceptibilities from both the initial
			and final states of the ZFC/FC relaxation at every $T_\mathrm{R}$,
			and checking whether they exhibit the anomalies of the III-IV phase transition
			or not.
			The details of the measurements are described in the Appendix.
			Hereafter, we describe the susceptibility measured from the initial and final states,
			$\mathrm{IV^{I/F}_{ZFC/FC}}(T_\mathrm{R})$, as a function of temperature and state,
			$\chi(T;X)$, $ X=\mathrm{IV^{I/F}_{ZFC/FC}}(T_\mathrm{R})$.

			The results are shown in 
			Fig. \ref{chiF}.
			The susceptibilities
			$\chi(T;  \mathrm{IV^{I}_{ZFC/FC} (2\ K)})$,
			are identical to those plotted in the top panel of 
			Fig. \ref{chidc}.			
			At low temperature, $\chi (T; \mathrm{IV^{F}_{ZFC/FC}}(T_\mathrm{R}))$ keep the same magnitude relationships as those at the final states,
			such as $\chi (T; \mathrm{IV^F_{ZFC}}(2\ \mathrm{K})) >	\chi (T; \mathrm{IV^F_{FC}}(2\ \mathrm{K}))$,
			With increasing temperature from $T_\mathrm{R}$,
			the susceptibilities 
			$\chi (T; \mathrm{IV^{F}_{ZFC/FC}}(T_\mathrm{R}))$ increase slightly and 
			show broad maxima at the 
			$T_\mathrm{R}$-dependent temperatures, and then, steeply decrease around 7 K. 
			All the susceptibilities merge above 10 K.
			Figure
			\ref{chiF} (b) shows the temperature derivatives of these susceptibilities,
			$\mathrm{d} \chi ( T; X)/\mathrm{d} T$.
			Independent of the states $X$, any derivative curve exhibits the dip anomaly at around $T_\mathrm{N4}$, which indicates 
			that the final states of the relaxation go through the III-IV phase transition as well as the initial states.
			Together with the fact that no relaxation is observed from the initial state at
			$H_\mathrm{R} =  18.25$ kOe and $T_\mathrm{R}=10$ K,
			$\mathrm{III^I_{ZFC/FC}\ (10\ K)}$ (Fig. \ref{III_relax} (a)), which is just above $T_\mathrm{N4}$,
			this result indicates that
			the long-time relaxation occurs inside the phase IV.
			Therefore, the long-time relaxation is the intrinsic and
			characteristic nature of this phase and not the relaxation from the metastable phase III to the stable phase IV.

			When looking into the dip temperatures of each 
			$\mathrm{d} \chi (T;X))/\mathrm{d}T$ in more detail,
			one can see a slight but systematic difference among them.
			Hereafter, let 
			$T\mathrm{^d_{ZFC/FC}}$ be the dip
			temperature of the curves of 
			$\mathrm{d} \chi (T; \mathrm{IV^{I/F}_{ZFC/FC}}(T_\mathrm{R}))/\mathrm{d}T$.
			As seen in the inset of Fig. 13 (b), which is the enlarged plot of 
			the temperature derivatives at around 
			$T\mathrm{^d_{ZFC/FC}}$,	
			those curves are almost identical to each other.
			However, the derivative curves  with different cooling conditions deviate from each other between 6 K and 8.5 K,
			and the cooling-process-dependent dip temperature
			$T\mathrm{^d_{ZFC} = 7\ K}$ and $ T\mathrm{^d_{FC} = 6.5\ K}$ are found.
			Here we should note that when cooling conditions are identical to each other, 
			the dip temperatures are almost the same, regardless of $T_\mathrm{R}$ and 
			whether the starting point of the susceptibility measurements, 
			$\mathrm{IV^{I/F}_{ZFC/FC}}$, is the initial or the final states of the relaxation.
			This indicates the system falls into the different initial states below the dip temperatures 
			depending on the cooling process and 
			the final states after long-time relaxations still differ depending on the cooling process. 
			Moreover, the states at different $T_\mathrm{R}$ after the same cooling process 
			are similar to each other and, thus, 
			the bifurcation of the state by the different cooling processes occurs 
			only at the dip temperatures, i.e. at $T_\mathrm{N4}$.

			\begin{figure*}[t]
				\centering
				\vspace{10pt}
				\includegraphics[width=\linewidth]{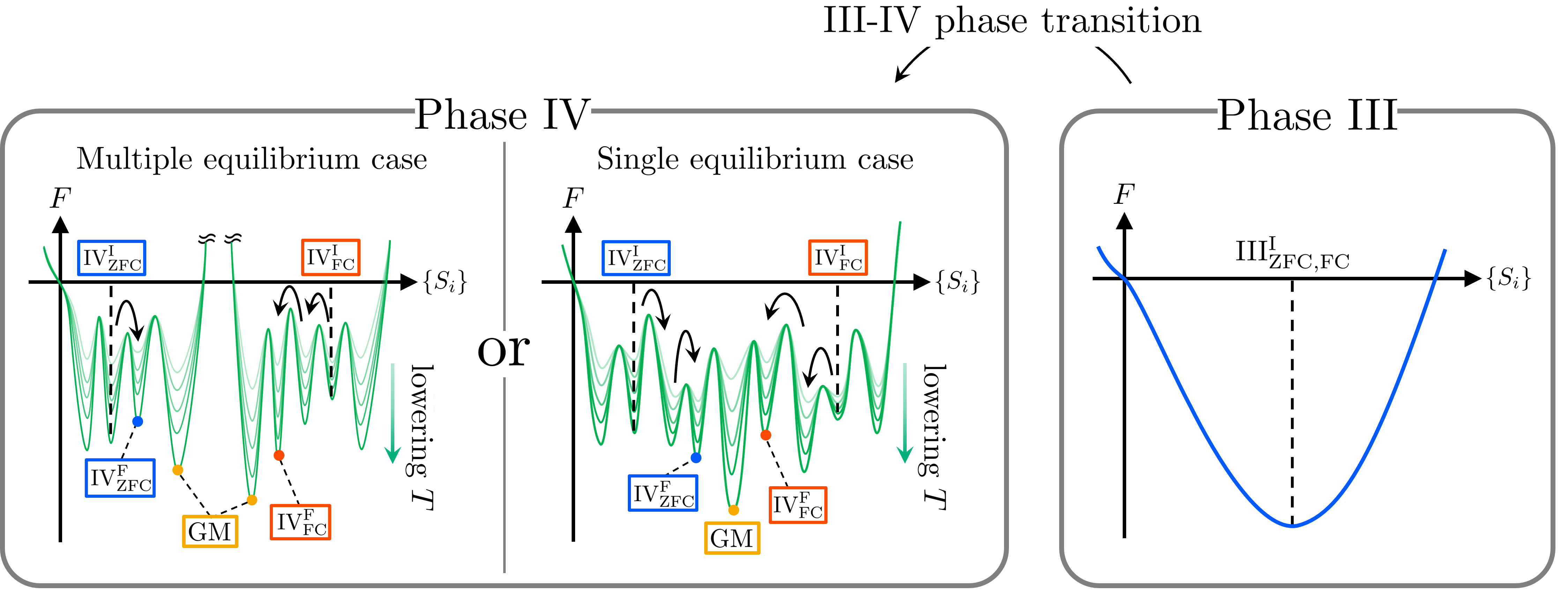}
				\caption{(Color online) 
							Schematics of the free-energy landscapes of the phases III (right) and IV (left).
							For the free-energy landscape of the phase IV, the left and right panels are the landscapes assuming the single and multiple equilibrium states, respectively. 
							The gradation of green curves represents the development of the depth of energy wells with temperature decrease.
							In the left panel, black arrows indicate the relaxations to the equilibrium state, where the different initial states reach the same equilibrium state (the global minimum, GM) on the right side 
							while, on the left side, they reach the different equilibrium states separated from each other with an infinite energy barrier in the thermodynamic limit.
							Note that the spin-reversal symmetry is already lost under the finite magnetic field and the translation symmetry breaking due to the long-period magnetic structure is not expressed in the figures for simplicity.
							}
				\label{free_energy}
			\end{figure*}
		
\ 
\section{Discussion}
\label{sec_discussion}

	\ \ 

	As a result of the measurements in the phase IV, 
	the following remarkable nonequilibrium phenomena were observed:

	\begin{itemize}
			\item The thermally-activated and long-time relaxations exceeding $10^5$ sec are observed in the phase IV.
			\item The long-time relaxation is observed from any initial state of the phase IV, 
			e.g., the plateau states prepared by the ZFC and FC processes and the states in the DF processes of both the whole and minor loops.
			\item Comparing the ZFC and FC relaxation at one $T_\mathrm{R}$, the ZFC relaxation is faster and overtakes the FC one along the way. This indicates that
			the relaxation process and its final state depend on the cooling condition.
			\item The III-IV phase transition is observed in all the susceptibilities, $\chi(T;X)$, measured from metastable states 
					$X=\mathrm{IV^{I/F}_{ZFC/FC}} (T_\mathrm{R})$. 
					The behavior of  $\chi(T;X)$ depends on the starting point of the measurement 
					(initial or final state of the relaxation) and the cooling condition
					(ZFC or FC), 
					whereas, exhibits almost identical behavior independent of 
					$T_\mathrm{R}$ if the two cooling conditions are the same.
	\end{itemize}
	
	The relaxation in the phase IV is one of the most important results found in this study.
	Long time relaxation in the phase with long-range order is usually observed along with the 1st-order phase transition,
	which accompanies the complicated domain dynamics due to the coexistence of the stable- and metastable-phase domains.
	\cite{JPSJ.85.034722,JOP.Conf.273.Sces.2010} 
	However, the relaxation in the phase IV should not be attributed to such a process.
	This is because, this relaxation occurs inside the phase IV and 
	should not be attributed to the dynamics along with the III-IV phase transition (Fig. \ref{chiF}).
	Also, while the relaxation along with the 1st-order phase transition is usually observed in the vicinity of the transition temperature or field,
	this is not the case for the relaxation in the phase IV.
	Rather a slight change in the magnetic structure is expected in this relaxation as observed inside the GH loop of 
	$\mathrm{Ca_3Co_2O_6}$ by neutron experiment\cite{JPSJ.80.034701.}.

	The series of relaxation measurements indicate important fact about the III-IV phase transition.
	It should be stressed again that 
	the most striking contrast between the phases III and IV is that 
	the long-time relaxation is observed in the latter, while it is absent in the former.
	It is indicated that, in the phase IV, the system searches the phase space for an extremely long time to reach the equilibrium while
	it is obtained as soon as the system transitions into the phase III.
	This suggests that the phase IV is the non-ergodic phase while the phase III is the ergodic phase
	and, thus, the III-IV phase transition is the ergodic to non-ergodic transition.
	This is highly nontrivial because the ergodic to non-ergodic transition is generally expected in random systems such as the spin glass,
	while this is not the case for the phase IV since it has the long-range magnetic order 
	\cite{PhysicaB.346-347.99}.
	
	Next, we qualitatively discuss the free-energy landscape in the phases III and IV.
	The phase III plateau is the thermal equilibrium state and, as shown in the phase diagram 
	(Fig. \ref{pha_dia}), the I-III and para-III phase transitions are either conventional first- (low temperature) or second-order one (high temperature).
	As reported earlier
	\cite{PhysicaB.346-347.99}, the magnetic structure of the phase III is the 9 times periodic one of the crystal structure along both the a- and b-axes.
	Therefore, a spontaneous break of the translational symmetry reflecting this period occurs with the phase transition into the phase III.
	The absence of the long-time relaxation indicates, 
	except for the structure corresponding to this symmetry-breaking, 
	the simple free-energy landscape in the phase III 
	(Fig. \ref{free_energy}, right panel).		
	As the system transitions into the phase IV, extremely long-time relaxation occurs.
	This indicates that the free-energy landscape in this phase 
	becomes complex with multiple local minima corresponding to metastable states, 
	which are separated from each other with quite high energy barriers (Fig. \ref{free_energy}, left panel).
	It is so complex that the relaxations from the initial states take a very long time to reach the global minimum(s) 
	which corresponds to the true equilibrium state(s) of the phase IV and it may be experimentally unreachable.

	The fact that the relaxation process and its final state depend on the cooling process in the phase IV 
	can also be explained based on such a complex free-energy landscape as follows.
	Assuming that the initial states are trapped in the different local minima depending on the cooling protocol, 
	the system traces the different paths on the free energy surface to approach the global minimum (equilibrium state).
	This can be observed as the difference in magnetization between ZFC and FC relaxations.
	The overtaking of the FC relaxation by the ZFC relaxation and 
	the distinct final states of the relaxations indicate 
	the distance between the initial states 
	$\mathrm{IV^I_{FC}}$ and $\mathrm{IV^I_{ZFC}}$ is so large in the phase space that their relaxation paths do not meet within the experimental time. 
	Otherwise,
	the ZFC and FC relaxations should merge and 
	do not intersect as observed in the present experiments.
	
	The systematic difference and similarity of 
	$\chi(T;X)$ (Fig. \ref{chiF}) can be interpreted using our picture.
	The difference can be interpreted as that 
	the positions of the initial and final states of relaxations in the phase space differ depending on the cooling condition, ZFC or FC. 
	However, this interpretation is not enough to understand the similarity, that is, when comparing the final states of the relaxations for the identical cooling conditions, 
	the susceptibilities measured from these states are similar to each other even if $T_\mathrm{R}$ differs.
	Here, for the free-energy landscape, we assume that the change in temperature 
	mostly contributes to the development of the depth of each energy well,
	while its position is almost invariant.
	This temperature change is schematically drawn in the left panel of Fig. \ref{free_energy}.
	If so, in the phase space, the positions of the initial states, the relaxation paths, and 
	the local minima in which the final states are trapped are almost 
	independent of $T_\mathrm{R}$ when the cooling condition is identical.
	Reflecting the differences or similarities of the final states as described so far, 
	when the temperature is increased from each final state and the susceptibility is measured, 
	its temperature dependence mainly depends on the cooling condition, ZFC or FC, but not on $T_\mathrm{R}$.

	The complex free-energy landscape in the phase IV should exhibit not only the temperature variation but also the time evolution, which has been discussed in spin glasses and structural glasses 
	\cite{ODAGAKI2021119448}. 
	In our experiments, the observed cycle-dependent minor loop (Fig. \ref{minor_loop_wh}) and 
	cooling-condition-dependent dip temperature of the susceptibility
	$\chi (T; X)$ (Fig. \ref{chiF} (b)) can be interpreted as the manifestation of the time evolution of the free-energy landscape.  	
	In the minor loop hysteresis measurements, the magnetic field is swept across the phase IV within the GH loop. In this case, it is expected that the free energy landscape cannot change instantly when the magnetic field enters or leaves the phase IV region, and the change is thought to be delayed with respect to the magnetic field sweep. 
	If this delay is very large, which is expected to be the case, the measurement would be performed under a different free energy landscape for each cycle, and the cycle dependence will appear in the minor hysteresis loop. 
	In the experiment, this cycle-dependence disappeared when the cycle was repeated eight times (Fig. \ref{minor_loop_wh} (b)), 
	which may result from a synchronization between the cycle of the minor loop hysteresis measurement and the time change of the free energy landscape. 
	The change in the free-energy landscape is also expected to be delayed with respect to the temperature variation. 
	In this case,  the observed dip anomaly in the temperature dependence of the magnetic susceptibility, 
	$\chi(T;X)$, is not the anomaly of the thermal equilibrium III-IV phase transition, but corresponds to a nonequilibrium time change of the free-energy landscape. 
	As discussed above, under the different cooling conditions, the system is trapped at completely different positions in the phase space when entering into the phase IV, and the time variations of the complex free energy landscape at the different positions can be different from each other. 
	As a consequence, the cooling-process dependent dip temperatures are observed (Fig. \ref{chiF} (b)). 
	Presumably, the dip temperature would be identical if the measurements were made over a sufficiently long time, which is not achievable in practice. 
	
	In the phase IV, the relaxation paths in the different cooling processes 
	do not meet each other within the experimental time scale. 
	This fact raises the question of whether the relaxation paths finally merge after sufficiently long-time observation or 
	never meet even in the infinite-time limit, in other words, whether the equilibrium state is unique or multiple.
	This difference leads to an intrinsic difference in the structure of free energy,
	namely, "single-valley" or "multi-valley".
	If the equilibrium state is unique, there exists only one global minimum corresponding to the equilibrium state  
	separated from local minima by high but finite energy barriers in the thermodynamic limit (Fig. \ref{free_energy}, the right side of the left panel).
	If the true equilibrium states are multiple, 
	there exist multiple global minima corresponding to different equilibrium states with different magnetization values
	separated from each other by infinite energy barriers in the thermodynamic limit (Fig. \ref{free_energy}, the left side of the left panel).
	In this multiple-equilibrium-state scenario, the cooling-condition-dependent relaxation 
	can be considered as the relaxation from an initial state towards an equilibrium state
	in each valley which is a region separated by infinite energy barriers in the phase space.
	The change of the free-energy landscape from single-valley to multi-valley should occur along with the III-IV phase transition.
	In this case, the multiple equilibrium states have different magnetizations and 
	are not linked by the trivial symmetry of the hamiltonian and, thus, it is considered that
	the phase IV is not an ordered state with the trivial symmetry breaking such as the ferromagnetic, antiferromagnetic, or ferrimagnetic phase. 
	Instead, one can expect a breaking of an unconventional symmetry, such as the replica symmetry breaking in the spin glass
	\cite{PhysRevLett.43.1754}.
	As shown in 
    Fig. \ref{determination_of_eqstate}, the relaxations from the several initial states prepared by the different conditions 
	are extrapolated to the different infinite-time values, 
	which may indicate the multiple equilibrium states in the phase IV. 
	However, due to the limitation of the measurement time, it is difficult to conclude experimentally whether the equilibrium state is unique or multiple.
	Further inspection into this point is needed.
		
	The phenomenon where the overtaking of the FC relaxation by the ZFC relaxation resembles the Mpemba effect (ME)
	\cite{PhysEdu.4.1969}.
	The ME is the anomalous nonequilibrium relaxation phenomenon 
	in which a system starts to freeze faster in the case of cooling from a hotter initial state than from a colder initial state.
	It is reported in several systems such as liquids
	\cite{PhysRevLett.119.148001,kjce.33.1903.2016} and granular gases
	\cite{PhysRevE.99.060901,PhysRevE.102.012906} and 
	recently reported in the numerical simulation on the spin glass
	\cite{pnas.1819803116}.
	Some theoretical works have quantitatively analyzed the ME by the Markovian dynamics using the distance-from-equilibrium function, 
	$\mathcal{D}$
	\cite{PNAS2018,PhysRevX.9.021060}.		
	The function
	$\mathcal{D} (\bm{p}(t), T_\mathrm{b})$ is literally a measure of how ``far" a probability distribution
	$\bm{p}(t)$ at time
	$t$ is from the equilibrium distribution at a given temperature
	$T_\mathrm{b}$.
	With this function, the ME is interpreted as the overtaking of the distance during the relaxation, namely,
	the distance of the initially hotter system becomes shorter than that of the initially colder one during their relaxation processes at a certain temperature.
	It is usually difficult to extract such a distance function in experiments, however, in magnets, 
	the difference in magnetization from the equilibrium value should be a measured quantity of the distance from the equilibrium state.
	Under this hypothesis, our discovered overtaking phenomena of the FC relaxation by the ZFC relaxation in the phase IV 
	can be interpreted as a special case of the ME,
	which seems to be proper to be referred to as the ``magnetic ME".
	The FC magnetization is usually larger and closer to the equilibrium one than the ZFC magnetization.
	This means that the initial state prepared in the FC condition
	($\mathrm{IV^I_{FC}}$) has a shorter distance than 
	the initial state prepared by the ZFC condition
	($\mathrm{IV^I_{ZFC}}$).		
	That is, the relaxations in two cooling conditions, ZFC and FC, in our experimental setup 
	correspond to the coolings from the hotter and colder states, respectively, and
	the overtaking phenomenon in the magnetization relaxations corresponds to the overtaking of the distance in the ME.

	At last, we discuss the cause of the GH loop and its potential interests.
	It is indicated that the strong nonequilibrium effect in the phase IV causes 
	the magnetization in the DF process not to return to the phase IV plateau within the typical timescale of magnetic hysteresis measurements,
	leading to the formation of the GH loop at every temperature below $T_\mathrm{N4}$.
	Note that not every phase inside the GH loop is non-ergodic.
	In other words, the existence of such a non-ergodic phase is a sufficient condition of the appearance of the GH loop.
	In fact, 
	even though the phase III plateau is inside the GH loop at 2 K, it is the ergodic phase.
	In contrast to the phase IV, the phase III is outside the GH loop above 3.5 K (Fig. \ref{M-H})  and such a stable phase seems to have little to do with the formation of the GH loop.
	Even so, if one investigates inside the GH loop, they can expect to find the non-ergodic phase or interesting nonequilibrium phenomena 
	such as the ``magnetic ME" in other materials.

\section{Summary}

	We investigated the nonequilibrium phenomena of the phase III and IV inside the global hysteresis (GH) loop 
	of the frustrated magnet $\mathrm{DyRu_2Si_2}$ using a single crystal sample.
	Especially, the plateau of the phase IV exists only inside the GH loop.
	By relaxation measurements from these plateaus and a few other measurements, 
	it is made clear that the long-time relaxation is intrinsically observed in the phase IV while it is absent in the phase III. 
	The relaxation in the phase IV is extremely long, in the order of $10^5$ sec, and is observed from any initial state, 
	prepared by the ZFC and FC process, and the DF process in the whole and minor loops.
	Quite unconventionally, the relaxation from the initial state prepared in the ZFC condition 
	overtakes that from the initial state prepared in the FC condition.  
	This indicates that the relaxation process and its final state depend on the cooling condition.
	The relaxation is so long that the equilibrium state is not reachable within the experimental timescale.
	The extremely long-time relaxation in the phase IV indicates that 
	the phase transition into the phase IV is not a conventional type, but an ergodic to nonergodic phase transition.
	The free-energy landscape of the phase IV should be complicated where  
	high energy barriers separate multiple local minima.
	On the other hand, the landscape of the phase III should be conventionally simple because the long-time relaxation is absent.
	The III-IV phase transition accompanying this change in the structure of the free-energy landscape from simple to complex should be a quite novel case 
	because such an ergodic to non-ergodic transition is usually expected only in random glassy systems.
	On the other hand,
	$\mathrm{DyRu_2Si_2}$ has a regular crystal structure and is not a random magnet, 
	and further, the regular long-range ordered magnetic structure was reported in the phase IV.
	Such an unexpectedly strong nonequilibrium effect in 
	this material should originate from strong frustration effects due to the long-range RKKY interaction.
	To get more insight into those nonequilibrium phenomena, we plan to monitor the time evolution of the magnetic structure in the phase IV along the relaxation by neutron scattering.

\section*{Acknowledgement}

The authors acknowledge the support from JST SPRING, Grant Number JPMJSP2110, and the JSPS
Grant-in-Aid for Scientific Research (B) (No. 20H01852).
	
\section*{Appendix}
\setcounter{figure}{0}
\appendix
\renewcommand{\thefigure}{A.\arabic{figure}}

		\begin{figure}[t]
				\centering
				\includegraphics[width=0.95\columnwidth]{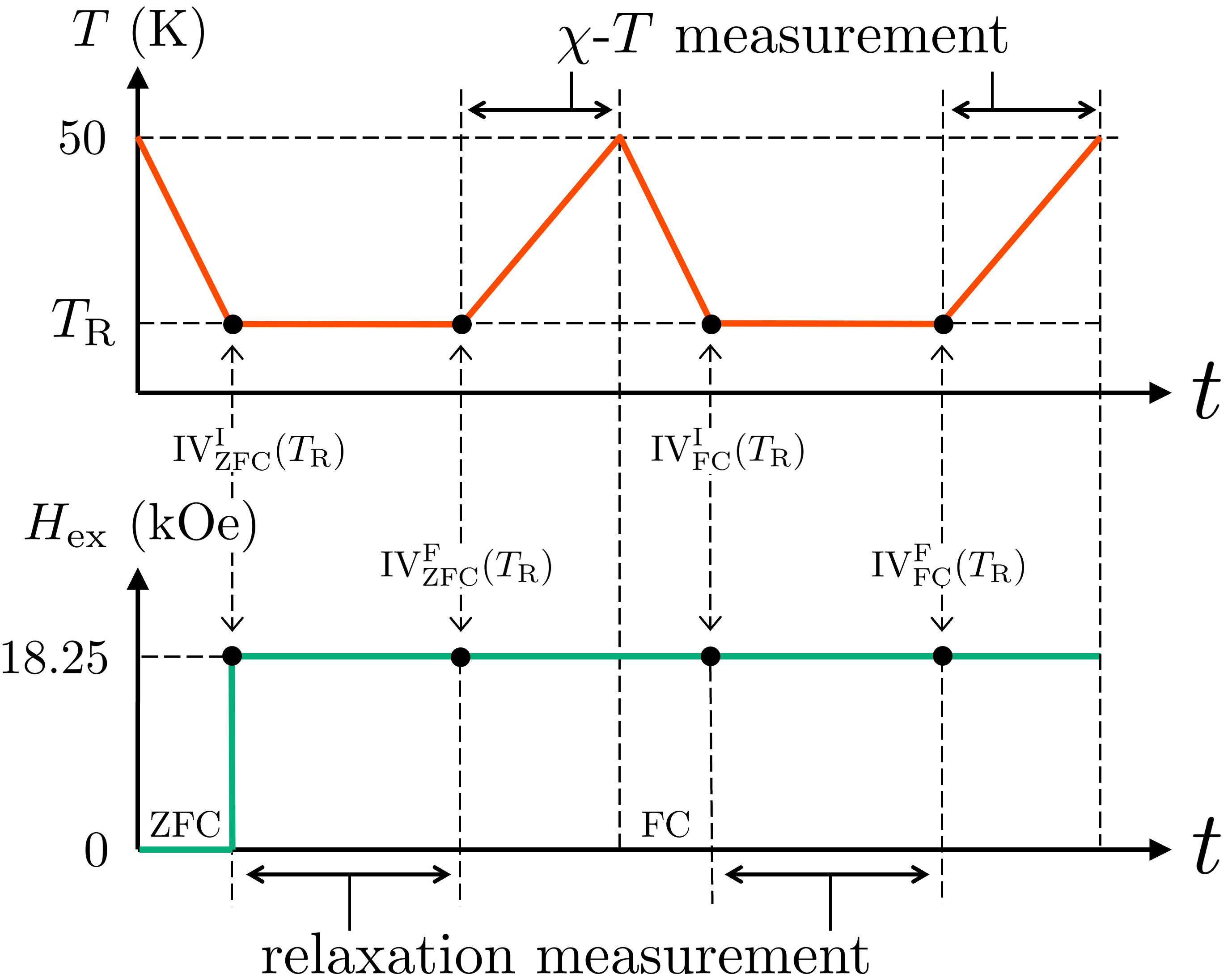}
				\caption{(Color online) 
					Profiles of the temperature and field changes for the preparation of the initial states,
					$\mathrm{IV_{ZFC/FC}^I}(T_\mathrm{R})$, at one $T_\mathrm{R}$ and
					the following measurements of the relaxation and the temperature dependence of susceptibility.  
						}
				\label{protocol}
		\end{figure}
	
	Here we explain the details of the magnetization relaxation measurements from the initial states $\mathrm{IV_{ZFC/FC}^I}(T_\mathrm{R})$.
	Generally, when measuring magnetizations of an identical sample with MPMS, the absolute value of magnetization would change whenever the sample is inserted into the machine for a new measurement.
	This is caused, for example, by the sample position against the pickup coil being not always precisely the same.
	For strict and quantitative discussion, the measurements should be performed in one run.
	Especially, this point is crucial when comparing the ZFC and FC relaxations. 
	
	For this purpose, the relaxations in two cooling conditions, ZFC and FC, at particular $T_\mathrm{R}$ and 
	susceptibilities from each final state of the relaxation, $\chi(T;X)$, were measured in one run.
	The profiles of temperature and magnetic field change in this measurement are shown in Fig. \ref{protocol}.
	In this measurement protocol, first the initial state 
	$\mathrm{IV^I_{ZFC}}(T_\mathrm{R})$ was prepared by cooling the sample down to
	$T_\mathrm{R}$ in the ZFC condition followed by applying $H_\mathrm{ex} = 18.25$ kOe.
	Then the relaxation was measured and, after that, from the final state of the relaxation,
	the temperature dependence of the susceptibility, 
	$\chi (T; \mathrm{IV^{F}_{ZFC}}(T_\mathrm{R}))$, up to 50 K was measured  (Fig. \ref{chiF}).
	Consecutively,  with maintaining the magnetic field, the initial state 
	$\mathrm{IV^I_{FC}}(T_\mathrm{R})$ was prepared by cooling the sample down to the same
	$T_\mathrm{R}$ in the FC condition.
	Similarly, the relaxation and the following temperature dependence of 
	$\chi(T;\mathrm{IV^F_{FC}}(T_\mathrm{R}))$ were measured.
	
	As the relaxations in the phase IV are extremely long, the measurements at distinct $T_\mathrm{R}$ unfortunately needed to be performed in the different runs
	and, thus, the absolute value of the magnetizations at different $T_\mathrm{R}$ can slightly differ.
	These values should be normalized as well.	
	Since the possible relaxation and memory effects due to the strong nonequilibrium effect at lower temperatures,
	which can depend on the measurement process, are erased in the paramagnetic phase,
	the susceptibility in the paramagnetic phase is inherently independent of the measurement process. 
	As a consequence, the susceptibilities measured in different runs should be equal to each other in the paramagnetic phase regardless of their processes.
	Thus, the normalization factors for the magnetizations in the different runs can be evaluated
	by equalizing the susceptibilities in the paramagnetic phase for every data as shown in Fig. \ref{chiF} (a).

\ \\
\noindent
*E-mail: yoshimoto.subaru.63z@st.kyoto-u.ac.jp
\bibliographystyle{jpsj}
\bibliography{C:/texlive/texmf-local/bibtex/bib/local/ref.bib,
C:/texlive/texmf-local/bibtex/bib/local/ref_mpemba.bib}

\begin{thebibliography}{10}

\bibitem{J.Phys.Cond.7.1889}
B.~Andreani, G.~L.~F. Fraga, A.~Garnier, D.~Gignoux, D.~Maurin, D.~Schmitt, and
  T.~Shigeoka: J. Phys.: Condens. Matter {\bfseries 7}, 1889 (1995).

\bibitem{PhysicaB.212.343}
A.~Garnier, D.~Gignoux, D.~Schmitt, and T.~Shigeoka: Physica B: Condensed
  Matter {\bfseries 212}, 343 (1995).

\bibitem{Phys.Rev.B.97.134425}
A.~F. Gubkin, L.~S. Wu, S.~E. Nikitin, A.~V. Suslov, A.~Podlesnyak,
  O.~Prokhnenko, K.~Proke\ifmmode~\check{s}\else \v{s}\fi{}, F.~Yokaichiya,
  L.~Keller, and N.~V. Baranov: Phys. Rev. B {\bfseries 97}, 134425 (2018).

\bibitem{JPSJ.66.3996}
H.~Kageyama, K.~Yoshimura, K.~Kosuge, M.~Azuma, M.~Takano, H.~Mitamura, and
  T.~Goto: J. Phys. Soc. Jpn. {\bfseries 66}, 3996 (1997).

\bibitem{JPSJ.80.034701.}
T.~Moyoshi and K.~Motoya: J. Phys. Soc. Jpn. {\bfseries 80}, 034701 (2011).

\bibitem{Phys.Rev.B.70.064424}
V.~Hardy, M.~R. Lees, O.~A. Petrenko, D.~M. Paul, D.~Flahaut, S.~H\'ebert, and
  A.~Maignan: Phys. Rev. B {\bfseries 70}, 064424 (2004).

\bibitem{PhysicaB.346-347.99}
S.~Kawano, M.~Takahashi, T.~Shigeoka, N.~Iwata, and M.~Bull: Physica B:
  Condensed Matter {\bfseries 346-347}, 99 (2004).

\bibitem{PhysEdu.4.1969}
E.~B. Mpemba and D.~G. Osborne: Phys. Edu. {\bfseries 4}, 172 (1969).

\bibitem{PhysRevLett.51.911}
L.~Lundgren, P.~Svedlindh, P.~Nordblad, and O.~Beckman: Phys. Rev. Lett.
  {\bfseries 51}, 911 (1983).

\bibitem{ODAGAKI2021119448}
T.~Odagaki, Y.~Saruyama, and T.~Ishida: J. Non-Cryst. Solids {\bfseries 558},
  119448 (2021).

\bibitem{PhysRevLett.43.1754}
G.~Parisi: Phys. Rev. Lett. {\bfseries 43}, 1754 (1979).

\bibitem{PhysRevLett.119.148001}
A.~Lasanta, F.~Vega~Reyes, A.~Prados, and A.~Santos: Phys. Rev. Lett.
  {\bfseries 119}, 148001 (2017).

\bibitem{kjce.33.1903.2016}
Y.~Ahn, H.~Kang, D.~Koh, and H.~Lee: Korean J. Chem. Eng. {\bfseries 33}, 1903
  (2016).

\bibitem{PhysRevE.99.060901}
A.~Torrente, M.~A. L\'opez-Casta\~no, A.~Lasanta, F.~V. Reyes, A.~Prados, and
  A.~Santos: Phys. Rev. E {\bfseries 99}, 060901 (2019).

\bibitem{PhysRevE.102.012906}
A.~Biswas, V.~V. Prasad, O.~Raz, and R.~Rajesh: Phys. Rev. E {\bfseries 102},
  012906 (2020).

\bibitem{pnas.1819803116}
M.~Baity-Jesi, E.~Calore, A.~Cruz, L.~A. Fernandez, J.~M. Gil-Narvi^^c3^^b3n,
  A.~Gordillo-Guerrero, D.~I^^c3^^b1iguez, A.~Lasanta, A.~Maiorano,
  E.~Marinari, V.~Martin-Mayor, J.~Moreno-Gordo, A.~M. Sudupe, D.~Navarro,
  G.~Parisi, S.~Perez-Gaviro, F.~Ricci-Tersenghi, J.~J. Ruiz-Lorenzo, S.~F.
  Schifano, B.~Seoane, A.~Taranc^^c3^^b3n, R.~Tripiccione, and D.~Yllanes:
  Proc. Natl. Acad. Sci. U.S.A. {\bfseries 116}, 15350 (2019).

\bibitem{PNAS2018}
Z.~Lu and O.~Raz: Proc. Natl. Acad. Sci. U.S.A. {\bfseries 114}, 5083
  (2017).

\bibitem{PhysRevX.9.021060}
I.~Klich, O.~Raz, O.~Hirschberg, and M.~Vucelja: Phys. Rev. X {\bfseries 9},
  021060 (2019).

\end{thebibliography}

\end{document}